# Comprehensive Study of Security and Privacy of Emerging Non-Volatile Memories

**Mohammad Nasim Imtiaz Khan, Member, IEEE, Swaroop Ghosh, Senior Member, IEEE**

School of EECS, The Pennsylvania State University, University Park, PA 16801 USA

Corresponding author: Mohammad Nasim Imtiaz Khan (e-mail: nasimimtiaz.khan@gmail.com).

The work is supported by the NSF under Award No. CNS - 1722557, CCF - 1718474, DGE - 1723687 and DGE - 1821766, DARPA Young Faculty Award under Award No. D15AP00089 and SRC under Award No. 2847.001.

**ABSTRACT** At the end of Silicon roadmap, keeping the leakage power in tolerable limit and bridging the bandwidth gap between processor and memory have become some of the biggest challenges. Several promising Non-Volatile Memories (NVMs) such as, Spin-Transfer Torque RAM (STTRAM), Magnetic RAM (MRAM), Phase Change Memory (PCM), Resistive RAM (RRAM) and Ferroelectric RAM (FeRAM) are being investigated to address the above issues since they offer high density and consumes zero leakage power. On one hand, the desirable properties of emerging NVMs make them suitable candidates for several applications including replacement of conventional memories. On the other hand, their unique characteristics such as, high and asymmetric read/write current and persistence bring new threats to data security and privacy. Some of these memories are already deployed in full systems and as discrete chips and are believed to become ubiquitous in future computing devices. Therefore, it is of utmost important to investigate their security and privacy issues. Note that these NVMs can be considered for cache, main memory or storage application. They are also suitable to implement in-memory computation which increases system throughput and eliminates Von-Neumann Bottleneck. Compute-capable NVMs impose new security and privacy challenges that are fundamentally different than their storage counterpart. This work identifies NVM vulnerabilities, attack vectors originating from device level all the way to circuits and systems considering both storage and compute applications. We also summarize the circuit/system level countermeasures to make the NVMs robust against security and privacy issues.

## I. INTRODUCTION

Conventional volatile memories such as, Static RAM (SRAM) and Dynamic RAM (DRAM) suffer from significant leakage power whereas conventional storage class Non-Volatile Memories (NVMs) such as, Flash memory suffer from higher write energy, poor performance, and endurance. However, emerging NVMs are beneficial due to their desirable properties - zero leakage, high-density, scalability and high endurance [1]. Some examples of emerging NVMs are Spin-Transfer Torque RAM (STTRAM) [2-5], Magnetic RAM (MRAM) [6-8], Phase Change Memory (PCM) [9-11], Resistive RAM (RRAM) [12-14] and Ferroelectric RAM (FeRAM) [15-16]. Emerging NVMs can also bridge the widening performance gap between processor and memory. Majority of these memories are also compatible with conventional Complementary Metal Oxide Semiconductor (CMOS) technology enabling easy integration with the logic process. Due to promising aspects, emerging NVMs are already being commercialized by industries e.g., Everspin (MRAM) [17], Adesto (RRAM) [18], Intel/Micron (PCM) [19] and Cypress (FeRAM) [20]. Intel's 3D Xpoint memory [19] is a recent example of NVM's adoption as a cache for Solid State Drives. Application of NVMs range from energy-harvested Internet-of-Things (IoT) and normally-OFF devices such as, body sensors, infrastructure health monitors [21] all the way to Supercomputers. Significant effort has been devoted to integrating emerging NVMs (note, henceforth the term 'NVM' is used to denote 'emerging NVM') in different levels of memory hierarchy [2-16, 21-44]. PCM and Resistive RAM (RRAM) offer high TMR but are limited by their endurance/power requirement and are mainly considered for main-memory applications [9-10, 14]. FeRAM is more suited towards main-memory applications due to a destructive read operation [16]. Spintronic memories like STTRAM and MRAM, offer very high endurance, high density and low voltage operation, and are therefore more suitable towards cache memory applications [2-4, 7-8].

Prior research has shown significant energy and performance gain with NVMs [2, 5-6, 8, 24, 31, 38, 41, 44]. It can provide desired memory bandwidth and reliability in high-performance computing, enable In-Memory Computing (IMC), instant-ON feature and energy-efficiency in mobiles, and power-efficiency in IoTs. Although NVMs can reap energy and performance benefits it may face new security issues that were not perceived before.

**Motivational example:** STTRAM contains Magnetic Tunnel Junction (MTJ) as the storage element (Fig. 1 (a)). The MTJ contains free and fixed magnetic layers. The magnetic orientation of the MTJ free layer can be toggled from parallel to anti-parallel (or vice versa) by injecting current from sourceline to bitline (or vice versa). The security challenges of STTRAM pertain to persistent data and its sensitivity to ambient parameters that can be exploited for low-cost tampering. Thus, addressing RRAM is sensitive to temperature and gases which can be used to tamper with the data. Broadly, there are two major threats to NVMs' security:

**(i) Threat to data security –** It pertains to data corruption (functional/timing) or destruction by a malicious attack with the intention to launch Denial-of-Service (DoS) attack. The fixed layer of STTRAM is robust, however, the free layer could be toggled using both spin-polarized current and magnetic field. Therefore, it is susceptible to manipulation through the magnitude and the polarity of the external magnetic field. Fig. 1 (b) and (c) show that the STTRAM free layer could flip its polarity either using current or with 250Oe magnetic field (easily produced by a horseshoe magnet [45]). Fig. 1 (d) shows the number of errors in commercial MRAM using a permanent magnet [46]. Similar results can also be obtained through temperature modulation. Ensuring data security against malicious attacks through ambient effect is particularly critical in deployed systems where enforcing and maintaining physical security are difficult.

**(ii) Threat to data privacy –** It pertains to the compromise of sensitive data (e.g., keys, passwords, credit card details) present in raw form through unauthorized access and side channels. The desire to have a larger Last Level Cache (LLC) for performance gain presents more persistent data that becomes vulnerable. Hard Disk Drive (HDD) has been the non-volatile part of the memory system. Encryption [47] is used to address the privacy of sensitive data of HDD. Volatile memory such as, SRAM is considered safe due to the randomization of data at power down. As non-volatility is introduced at higher levels of memory stacks, that were traditionally volatile, more data become vulnerable that were originally safe. As the memory level moves closer to the Central Processing Unit (CPU) it becomes more sensitive to latency. Consequently, the application of encryption in LLC is difficult. Thus, addressing data privacy in higher levels of memory stack while maintaining performance is a challenge. New measures are required to resolve this. Designing a magnetic or heat shield around the device can be a possible solution but owing to its cost and weight, it may not be practical or effective for a range of applications including IoTs and mobiles. Existing packaging specification only considers the ambient stray magnetic field (~25 Oe) [48]. However, NVMs can face an intentional magnetic field and temperature which may not be protected by the packaging itself. Everspin lists the maximum magnetic tolerance for their MRAM chips during write/read/standby to be only ~100Oe [17]. High and asymmetric write current [49], and, long and asymmetric write latency [50] (common in most NVMs) can serve as side channels exposing the number of '1's and '0's in a memory word weakening the data privacy [50-54]. Higher write current provides knob to the adversary to launch fault injection [55], information leakage [56] and row hammer attack [57] through voltage droop/ground bounce. Furthermore, the inherent non-volatility of NVMs can be leveraged to design NVM Trojan [72] which is difficult to detect during the testing phase or prevent from being activated using system-level mitigation techniques.

Interestingly, the implementations and usages of NVMs for various applications such as, cache, main memory, storage, and IMC vary vastly. For example, the cache may employ 1T-1S (S: storage), main memory and storage may use 1D-1S (D: diode) and, IMC may use either of them but with conceptually different read/write/computing modes. Therefore, the same attack model might not be applicable to leverage their vulnerabilities in other application modes. Furthermore, if they suffer from the same vulnerability, a common mitigation technique might not be applicable to all modes of application. For example, advanced wear-leveling techniques (with high computational overhead) might be suitable for the main memory application but not for cache memory due to tighter performance requirements. NVM Trojan triggers and payloads also differ with respect to NVM's placement in the cache hierarchy. For example, a Trojan trigger inside the processor can directly tap the raw addressing of L1 cache for sensitization, however, a trigger in the main memory has to work harder to extract the trigger signal (naturally obfuscated due to address translation). Therefore, NVM vulnerabilities, data security, and privacy issues should be investigated across all modes of applications for a deeper understanding of the challenges for the development of strong countermeasures.

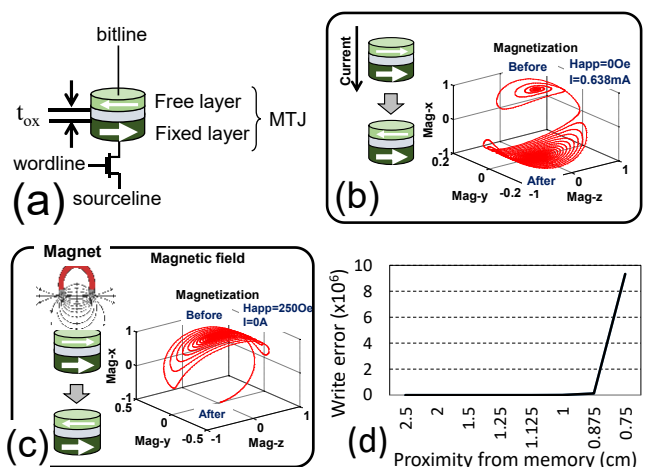

**FIGURE 1.** (a) Schematic of STTRAM; (b) flipping of MTJ free layer due to STT (Happ=0Oe, I=0.638mA); (c) flipping due to external magnetic field (Happ=260Oe, I=0). Plots obtained using LLG equation with 60x120x3nm MTJ, α=0.01, Δ=56, Ms=1000A/m, P=0.8; and, (d) write errors with magnetic attack strength obtained from commercial MRAM chip [45].

This paper is related to [104] that reviews the security properties and applications of spintronic devices. The detailed differences are as follows: (i) we present security and privacy issues of a broad range of emerging Non-Volatile Memories (NVMs) (such as, STTRAM, RRAM, MRAM, PCM and FeRAM) instead of focusing on only spintronic memories; (ii) we cover various vulnerability of various emerging NVMs such as, supply noise and data signature leakage etc. instead of vulnerabilities of spintronic memories such as, susceptibility to external magnetic field and persistence; (iii) we cover wide range of attack models such as, side channel attack, fault injection attack, information leakage attack, row hammer attack and Denial of Service (DoS) attack instead of covering only external magnetic field and probing after power down; (iv) we present security issues of NVM as memory (for cache, main memory and storage) as well as compute application.

The rest of the paper is organized as follows: Section II provides background on various NVM technologies and describes their vulnerabilities; Section III presents the privacy issues and countermeasures of NVM-based cache; Section IV describes the NVM enabled hardware Trojan attacks; Section V explains the security issues and countermeasures of NVM-based cache; Section VI presents security and privacy analysis of NVM-based main memory and storage memory; Section VII summarizes the threats on the compute capable NVMs; Section VIII describes the test techniques to detect the security issues of NVMs after manufacturing; Section IX presents the future directions for NVM security and privacy research. Finally, Section X draws the conclusion.

## II. NVM DEVICES AND THEIR VULNERABILITIES

In this section, the basics of a few emerging NVM technologies are described along with their vulnerabilities. We mainly investigate STTRAM, MRAM and RRAM for the sake of brevity and use them for drawing general conclusions on emerging NVMs. We also introduce other flavors of NVMs as necessary.

**STTRAM/MRAM:** STTRAM bitcell (Fig. 2(a)) contains MTJ as the storage element which contains a free (FL) magnetic layer, a pinned (PL) (also known as fixed) magnetic layer and an oxide layer between them. Each bitcell also contains an n-MOS as the selector device. The resistance of the MTJ stack is high (denoted by RAP) if FL magnetic orientation is Anti-Parallel (AP) compared to the PL and the resistance is low (denoted by RP) if FL magnetic orientation is Parallel (P) compared to the PL. MTJ can be toggled from P state (data '0') to AP state (data '1') (or vice versa) using current-induced Spin-Transfer Torque by passing the appropriate write current (> critical current) from Sourceline (SL) to Bitline (BL) (or vice versa). MRAM bitcell (Fig. 2(b)) is similar as STTRAM. However, MRAM write operation is magnetic field-driven. Current is passed through BL and Digitline (DL) with appropriate direction and magnitude which flips the magnetic orientation of FL and thereby, writes the data. During a read operation, a small voltage is applied and the resistance of the bitcell is sense out for both STTRAM and MRAM.

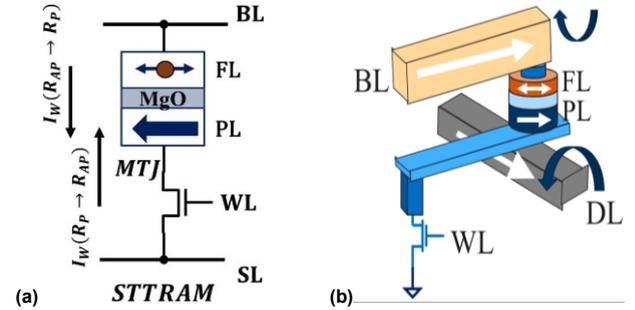

**FIGURE 2. Bitcell schematic of (a) STTRAM; (b) MRAM.**

**Vulnerabilities:** STTRAM/MRAM suffers from,
(i) High write current: This can lead to supply noise which can be leveraged in order to launch fault injection attack [54], DoS attack [54], information leakage attack [55], and row hammer attack [56];
(ii) Asymmetric [49] write and read current: This can be leveraged to launch side channel attack [50-54];
(iii) Susceptibility to external fields: External magnetic field can flip the magnetic orientation of FL of MTJ which corrupts the data [58]. It can be leveraged to launch DoS attack.
(iv) Susceptibility to temperature: High temperature can lead to reduced data retention and an adversary can trigger a DoS attack.

**RRAM:** RRAM contains an oxide material between two electrodes known as Top/Bottom Electrode (TE/BE) (Fig. 3).

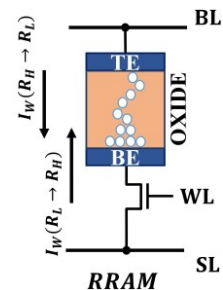

**FIGURE 3. Bitcell schematic of RRAM.**

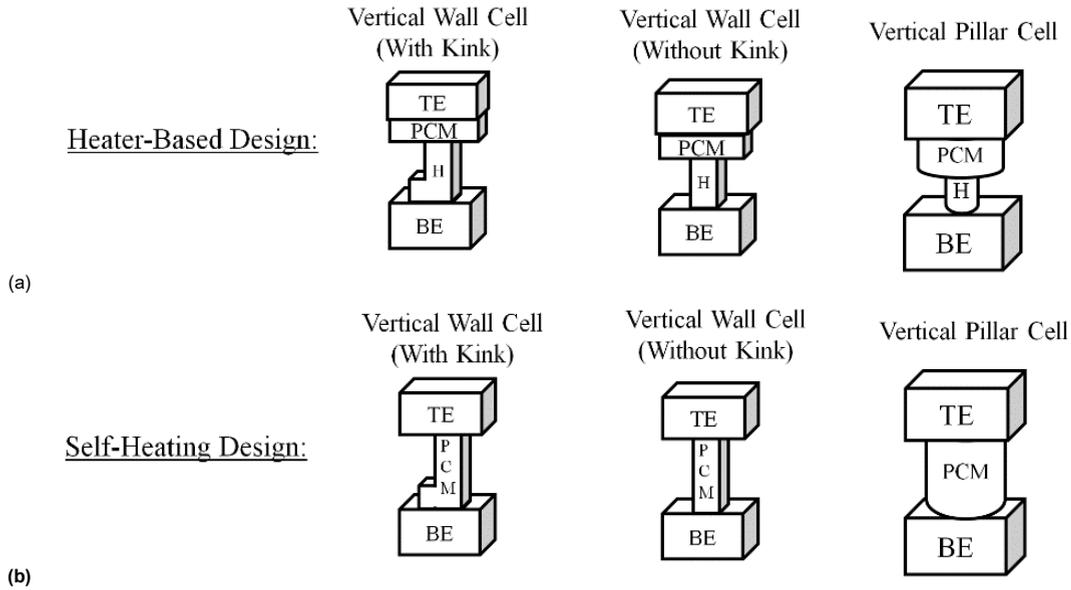

**FIGURE 4.** Some common design schemes for self-heating and heater-based PCM memory cell designs.

RRAM resistive switching is due to oxide break-down and re-oxidation which modifies a Conduction Filament (CF). Conduction through the CF is primarily due to the transportation of electrons in the oxygen vacancies. These vacancies are created under the influence of the electric field due to applied voltage. The two states of the RRAM are termed as Low Resistance State (LRS) and High Resistance State (HRS). The process of switching the state to LRS (HRS) is known as SET (RESET). During read operation, a small voltage is applied and the resistance of the bitcell is sensed out.

**Vulnerabilities:** RRAM also suffers from, (i) high write current; (ii) asymmetric write and read current; (iii) susceptibility to temperature; and, (iv) low endurance. The adversary can hammer a cell to exhaust the lifetime of RRAM.

**PCM:** PCM memory cells can be constructed in several different ways each PCM with its own advantages and disadvantages. PCM cell designs can be broken down into two primary categories: heater-based and self-heating-based [59]. Heater-based (Fig. 4 (a)) PCM memory cells rely on the existence of a locally placed layer of material to act as a heat reservoir and to heat the adjacent layer of phase-change material [59]. This is commonly done with materials such as, Tungsten or Titanium Nitride (TiN) [59-61]. Self-heating (Fig. 4 (b)) designs remove the heat reservoir from the design and instead rely on the internally generated heat within the PCM to cause the material to change the state [59]. Both designs typically rely on commonly used phase-change materials such as, Ge2Sb2Te5 (GST) [60, 62], however other variants exist that use other PCMs such as, In3Sb1Te2 [63-64]. Two of the most common types of PCM cell designs are the vertical wall and vertical pillar designs [59]. The device layer of the vertical pillar design is exactly like the vertical wall's design, only the geometry is that of a cylinder instead of a rectangular prism. For heater-based PCM memory cells, it is common to have the heating element be smaller than the PCM layer in order to localize the heating process to a particular location within the GST [60]. Other designs for PCM cells such as the planar bridge design also exist.

Writing to a PCM memory cell is the act of applying a current through the device that causes enough localized heating in order to force the device to either SET or RESET. The SET state within a PCM memory cell is typically a low resistance state where the structure of the phase-change material is crystalline. The SET state is usually achieved by keeping the PCM device above its melting temperature (Tmelt) for a long enough period of time to allow the device to enter the crystalline state. If the device is instead quickly quenched by quickly removing the current through the device during a shorter heating process, the crystal structure inside the device instead becomes amorphous. This process places the device in its RESET state. This causes the internal resistance of the device to be a much higher value, typically at least one order of magnitude higher or more [59]. During a read operation, a small voltage is applied and the resistance of the bitcell is sensed out.

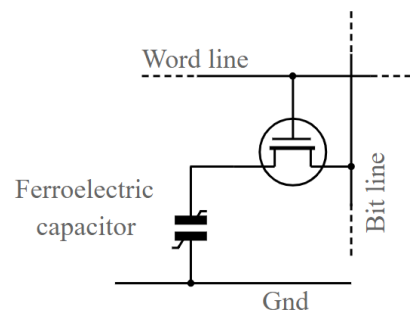

**FIGURE 5.** Bitcell schematic of FeRAM bitcell.

TABLE I
COMPARATIVE ANALYSIS OF SUSCEPTIBILITY OF DIFFERENT NVMs TO DIFFERENT DATA PRIVACY AND DATA SECURITY ATTACKS

| | Susceptible to | | | | | | | |
|---|---|---|---|---|---|---|---|---|
| | Data Privacy | | | | Data Security | | | |
| | Side Channel Attack | Fault Injection Attack | Information leakage Attack | Row Hammer Attack | DoS Attack | Thermal Attack | Magnetic Attack | Electric Field Attack |
| **STTRAM** | ✓ | ✓ | ✓ | ✓ | ✓ | ✓ | ✓ | |
| **MRAM** | ✓ | ✓ | ✓ | ✓ | ✓ | ✓ | ✓ | |
| **RRAM** | ✓ | ✓ | ✓ | ✓ | ✓ | ✓ | | |
| **PCM** | ✓ | ✓ | ✓ | ✓ | ✓ | ✓ | | |
| **FeRAM** | ✓ | | | | ✓ | ✓ | | ✓ |

**Vulnerabilities:** PCM suffers from, (i) high write current; (ii) asymmetric write and read current; and, (iii) low endurance. The adversary can hammer a cell to exhaust the lifetime of the PCM.

**FeRAM:** FeRAM bitcell (Fig. 5) contains one capacitor and one access transistor. The bitcell is similar to DRAM, except a dielectric structure containing ferroelectric material is used in FeRAM instead of a linear capacitor. During a write operation, an applied electric field across the ferroelectric layer is applied which forces the atoms inside into the 'up' or 'down' orientation (based on electric field polarity), thereby storing a '1' or '0'. During read operation, the transistor forces the cell into a particular state, say '0'. If the cell is holding a '0', bitline voltage remains the same. Otherwise, the re-orientation of the atoms in the film will cause a brief pulse of current in the output which indicates the cell held a '1'. Note that, FeRAM read is destructive in nature and requires the cell to be re-written.

**Vulnerabilities:** FeRAM suffers from, (i) high write current, (ii) asymmetric read current; and, (iii) susceptibility to an external electric field which can flip the polarization of ferroelectric material of FeRAM bitcell. The adversary can launch a DoS attack by applying an external electric field. FeRAM is also susceptible to ambient temperature.

### *B. COMPARATIVE ANALYSIS*

Table I summarizes the vulnerabilities of some of the emerging NVMs. The vulnerabilities can be categorized into data privacy and data security. Side channel, fault injection, information leakage, and row hammer attacks can be termed as data privacy attacks whereas DoS, thermal, magnetic and electric field attacks can be termed as data security attacks. From the table, it can be noted that spintronic memories are more vulnerable to the attacks while FeRAM is the least. In the following sections, the data security and privacy attacks on NVMs are explained.

### III. PRIVACY ISSUES AND COUNTERMEASURES OF NVM-BASED CACHE

In this section, the privacy issues related to the NVM-based cache and various countermeasures are explained.

### *A. SIDE CHANNEL ATTACK (SCA)*

In this subsection, NVM susceptibility to SCA is discussed taking STTRAM as a test case. However, other NVMs also exhibit similar susceptibility.

#### 1) BACKGROUND

SCA [65] is a powerful threat, which targets weak implementation of cryptographic algorithms rather than the algorithm itself. The implementation weakness is related to device physics of the underlying compute element, which makes it hard to fix the vulnerability. In general, SCA exploits the unintentional leakage observed in physical channels like timing [66], power consumption [65], or electromagnetic (EM) emanation [67] etc. with an objective to retrieve the secret key. Compared to conventional SCA that exploits the computing elements like SBOX to extract information, SCA on memory is based on the observation that read/write of sensitive data on the memory leaks information. Since different data transitions ($0\rightarrow1$ or $1\rightarrow0$) have different physical signatures (power consumption, write/read time and current), a physical signature dependent on secret key value can reveal the secret. The attack targets a key-related sensitive computation being performed on the target device under some leakage model assumptions.

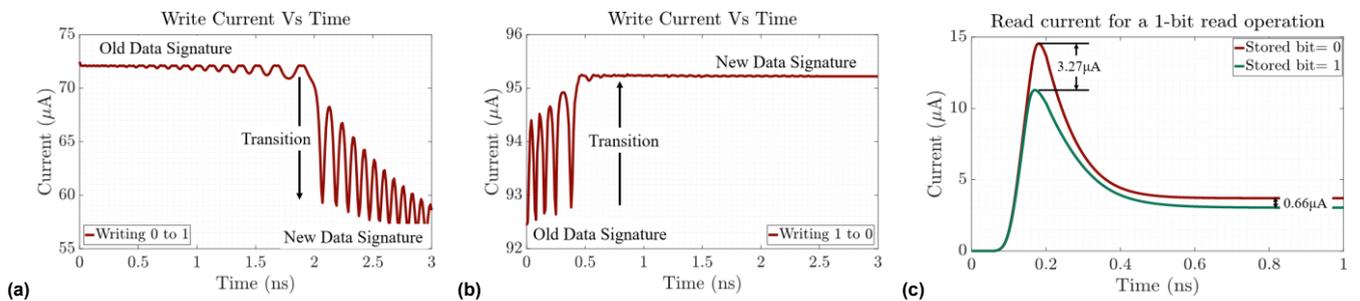

**FIGURE 6.** Supply current waveform for, (a) writing '0' to '1'; (b) writing '0' to '1'; and, (c) reading data '0' and '1'. The magnitude and waveform of write current is a function of stored data which also acts as a signature. Furthermore, a significant gap is present between write current of data '0' and '1' as well as read current for data '0' and '1' which can be leveraged as signature.

For memory side channel attack, Hamming Distance (HD) model is used, which equals the number of bit transitions, when the value in the memory cell is updated. Next, a statistical dependency is tested between the observed leakage (or side-channel measurements like power traces) and the hypothetical leakage computed using the leakage model and some hypotheses on the secret key.

One advantage of SCA over traditional cryptanalysis is that SCA can apply the divide and conquer approach. Thus, instead of testing and recovering the full key at once, SCA can be used to recover small parts of the key independently and later combined. Dividing the key into small parts allows to exhaustively test all hypotheses on the keys. For example, it is practically impossible to test 128-bits of the key in AES since there are $2^{128}$ possible combination. Instead, SCA can be applied on 8-bits of the key independently, which limits the key hypothesis to $2^8$. The rest of the 120 bit is considered as noise. For a successful attack, the correct key hypothesis shows a much higher statistical dependency as compared to all other key hypotheses. Once the first 8 bit of the key is recovered, the attack can be repeated 16 more times to retrieve the rest of the key. This reduces the complexity to $2^8$ x 16 = $2^{12}$ from $2^{128}$. Commonly used statistical tools are the Difference of Means (DoM) [67] and the Pearson correlation coefficient [68].

### 2) SCA ON STTRAM

The write current of STTRAM is dependent on the polarity of the stored data. MTJ equivalent resistance is high (low) in state '1' ('0'). Fig. 6(a) shows the supply current waveform for single-bit write '1' when the previously stored value is '0'.

Initially, the current is high (MTJ resistance low) but it reduces after successful write. Fig. 6(b) shows the supply current waveform for writing '0' with the previous value stored as '1'. In this case, the current is initially low and goes high after successful write. Therefore, write current waveform of $0\rightarrow1$ and $1\rightarrow0$ can be divided into three phases; 'old data', 'transition from old data to new data' and finally, the 'new data'. Write current waveform of $0\rightarrow0$ and $1\rightarrow1$ are relatively constant as the state is unchanged. The high and low states of the current (for writing $0\rightarrow1$ and $1\rightarrow0$) are very distinct and they reveal the information about the previous and new data. The current difference between the states depends on the Tunnel Magneto Resistance (TMR) [49] of MTJ. For a robust read operation, it is desired to have higher TMR which adversely affects the data privacy.

The read current for both data '0' and '1' (Fig. 6(c)) depends on the current state of the bit. Therefore, the key could be extracted when it is being read during the intermediate steps of an encryption operation (such as, AES or MICKEY-128 2.0). Furthermore, read/write operations can be distinctly identified from the corresponding current waveforms.

In [54], a Differential Power Analysis (DPA) on STTRAM read/write operation and MRAM read operation are performed. The work used the HD leakage model with Pearson correlation and attacked the last round of AES-128 encryption to investigate SCA vulnerability. The first byte of the key can be retrieved in around 600 traces which is suboptimal. The work further improved the attack model with some basic pre-processing (subtracting the average initial write current from the final write current, which results in the change in write current). The result is shown in Fig. 7(a). The black line is the correct hypothesis and the attack is considered successful when it emerges from the cloud of all wrong hypotheses. Similar analysis is also performed on STTRAM read operation and SRAM write operation to compare the results. Fig. 7(b) summarizes the result with the number of key bytes retrieved with respect to the required traces for STTRAM and SRAM. Before the pre-processing, STTRAM write was revealing 8 byte of the key in around 2000 traces whereas after pre-processing, STTRAM write reveals all 16 byte in around 1600 traces. Note that the later is similar to SRAM write operation. However, STTRAM read operation is more vulnerable since it leaks all 16 byte in just 400 traces.

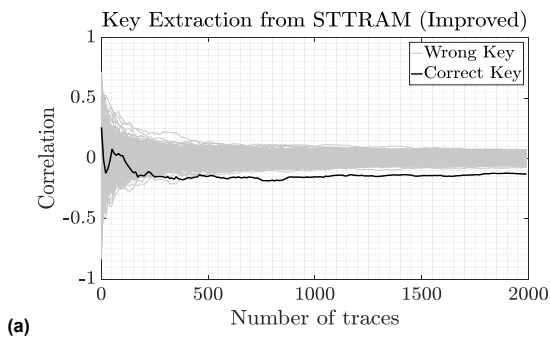

(a)

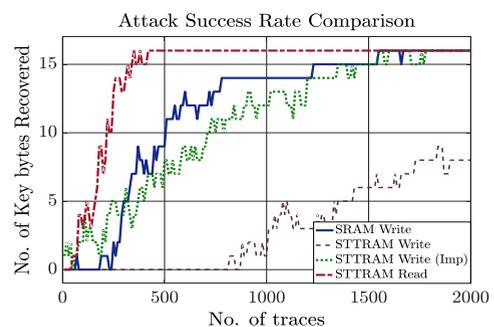

(b)

**FIGURE 7.** (a) Improved extraction from STTRAM after pre-processing; (b) comparison of the number of key bytes retrieved with respect to the number of traces for SRAM, STTRAM and STTRAM improved.

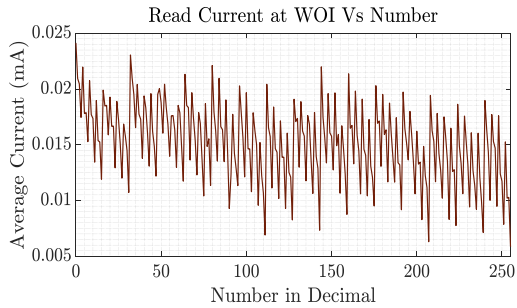

FIGURE 8. Average current at WOI vs number being read from chip.

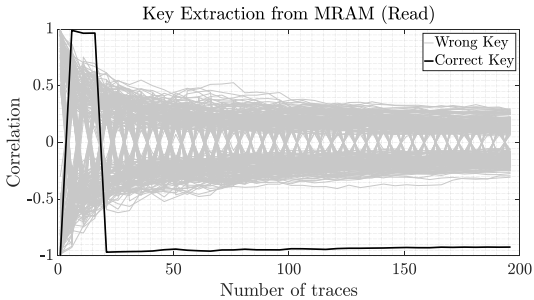

FIGURE 9. Correlation vs number of traces for MRAM read operation.

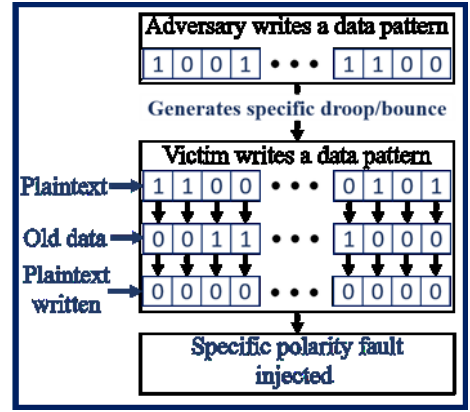

FIGURE 10. Specific polarity fault injection attack.

The work further verified their analysis on the read operation of a commercial MRAM chip. It has been identified that a window of 15ns i.e., Window of Interest (WOI) performs sensing of actual data from the memory cells. In Fig. 8, the average current in the WOI for reading data 0 to 255 is plotted. It is evident that the average read current depends on the number of ones in 8-bit read data. The trend that the read current reduces with the increasing Hamming Weight (HW) of the 8-bit data, proves that the asymmetric read current of MRAM reveals the sensitive data. The work further shows that the attack on MRAM read can retrieve the key (Fig. 9).

In [53], a Correlation Power Analysis (CPA) on MRAM write operation has been performed. The work proposes a hypothetical power model that considers the difference of $0 \rightarrow 1$ and $1 \rightarrow 0$ transitions to estimate the post-alignment power consumption while writing to MRAM. They considered the stream cipher MICKEY-128 2.0 to validate the proposed attack model. The results show that the secret key can be retrieved from MRAM write operation traces.

### 3) CONSIDERATIONS FOR OTHER NVMS

SCA exploits asymmetric write current to extract data. Since all NVMs (except technologies such as, Spin-Orbit Torque MRAM) incur asymmetric write and/or read current, they are equally susceptible to SCA.

### 4) COUNTERMEASURES

To weaken SCA, the data-dependent asymmetry in the read/write current should be eliminated or minimized. One possible technique is complementary bit encoding. For example, if the data bus is 8 bits, the first 4 bits can be written as original data while the last 4 bits can be written as their complement. While reading, the last four bits need to be complemented to recover the original data. This will require modification in the read/write circuitry; however, this will obfuscate the read/write data signature. It is still possible that the adversary might be able to extract the data, but the key extraction effort will be much more extensive. It is important to note that, although data encoding can be a potential countermeasure against SCA [66], it can be very sensitive to process variation. Thus, careful analysis is required before deploying such protection. Another approach to eliminating write asymmetry is to perform a constant current write [69]. In this method, instead of providing a constant voltage across the bitcell, a constant current will be passed through the cell with appropriate polarity to write the bit. However, this results in write power overhead.

### B. FAULT INJECTION ATTACK

In this subsection, NVM susceptibility to fault injection attack is discussed taking RRAM as a test case. However, other NVMs also exhibit similar susceptibility.

### 1) BACKGROUND

Fault injection attacks during write (read) operation can prevent successful writing of specific data polarity (read incorrect data for specific data polarity). Such attacks can be exploited to leak system assets such as, cryptographic keys. One example is when an adversary induces single-bit or multi-bit faults in a cryptographic system and performs differential fault analysis by observing correct and faulty pairs of inputs and outputs and subsequently derives simplified equations to extract the keys. Multiple methods for extracting keys using fault injection have been extensively studied and demonstrated [70]. In contrast to existing methods that focus on fault injection in clock and/or power rail, we describe techniques that inject fault in NVMs.

In [55], a fault injection attack on NVM (taking RRAM as a test case) using supply noise has been investigated. Fig. 10 shows a high-level idea. Adversary writes in his memory space and generates a specific supply noise which can propagate to a victim's memory space and cause failures to specific polarity read/write operation.

## 2) ATTACK MODEL AND ASSUMPTION

It has been shown [55] that an adversary can leverage high and asymmetric write current and long latency of NVMs (which causes supply voltage droop and ground bounce) to launch fault injection attack. An adversary can write specific data patterns (i.e., a specific number of 0's and 1's) to generate deterministic droop/bounce. This will propagate to the user's memory space and create read/write failure. For this analysis, an RRAM-based (i.e., 1T1R) Last Level Cache (LLC) has been considered as a test case.

The work [55] assumes that: i) NVM LLC is being shared by two users (i.e., an adversary and a victim); ii) bank-level parallel read/write operation is performed to increase the throughput; iii) adversary has the knowledge of the amount of droop/bounce that can be generated by a read/write data pattern; iv) adversary also knows how the generated droop/bounce propagates (decays with distance) and how it affects the victim's write/read operation; v) the adversary is an expert in computer architecture and can exploit knobs e.g., accessing specific data pattern in pre-defined physical locations to prevent their replacement by policies e.g., Least Recently Used (LRU).

## 3) SUPPLY NOISE DUE TO HIGH WRITE CURRENT

High current (50-100mA assuming ~100μA/bit) is drawn from the supply for a full cache line (512-1024bit) write. This creates two types of supply noise:

• Supply voltage droop: On-chip voltage regulator or power supply keeps the supply voltage constant. However, the supply voltage (distributed in metal M8) reaches the memory bitcell (implemented in metal M1) via the power-grid RC network. The interconnect resistance causes a significant voltage droop at the bitcell due to high current. Voltage droop lowers headroom for the bitcell and increases the write latency or decreases the sense margin for the read. It can eventually lead to a read/write failure.

• Local ground bounce: Similar to supply voltage, the true ground is routed on the upper metal layer (e.g. M8) and connects to the transistors in M1. Therefore, the local ground rail bounces when the charge (due to high write/read current) is dumped.

Droop/bounce magnitude depends on the present state of the memory bit as well as the new data being written since Iwrite for 0→0, 0→1, 1→0 and 1→1 is different (for a write operation), and on the stored data (for a read operation). For example, Fig. 11(a) shows the bounce generated by a full cache line write for various data patterns. It is notable that 1→1 write creates lowest, and 0→0 creates the highest bounce.

## 4) FAULT INJECTION USING WRITE OPERATION

RRAM write operation is simulated with additional supply noise generated by a parallel operation in another independent bank [56]. As supply noise increases, write latency for both LRS to HRS and HRS to LRS increases. Fig. 11(b) shows the RRAM resistance switching during write operation with respect to supply noise. Supply noise beyond 50mV and less than 120mV will cause LRS to HRS write failure but still can write HRS to LRS successfully. If the adversary can generate supply noise in a way that the victim incurs noise in this range, it will launch a 0→1 polarity fault injection attack. However, if the victim incurs combined voltage loss > 120mV, it will cause complete write failure i.e. DoS attack.

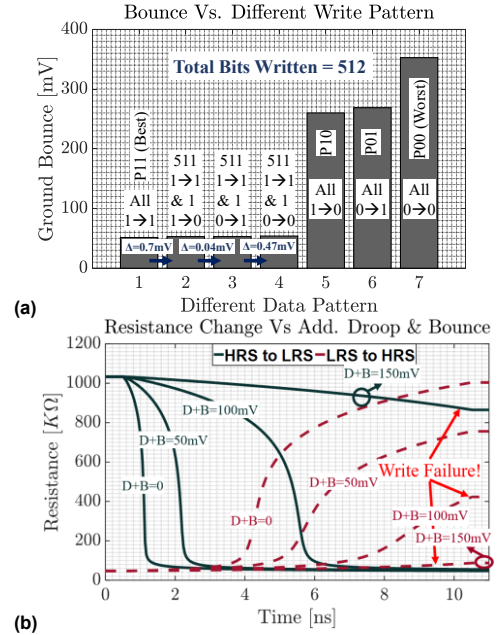

**FIGURE 11.** RRAM, (a) resistance variation with additional combined voltage loss; (b) latency increases as additional supply noise increases.

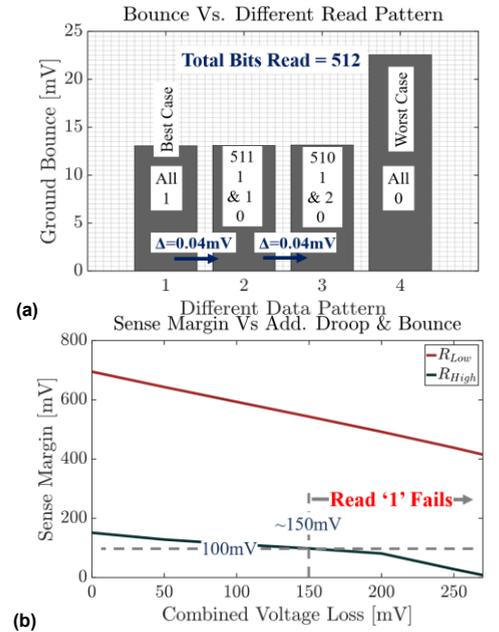

**FIGURE 12.** (a) Bounce generation vs different read data pattern; (b) sense margin with additional droop and bounce. Sense margin for data '1' suffers more, and failure is observed above 150mV of additional droop and bounce.

Detection of victim's write initiation: Adversary needs to know when and where the user is writing in order to launch DoS/fault injection/information leakage attack (discussed in the next subsection). One possible approach adopted by the adversary is to store data that generates high noise (i.e., all 0) in various locations of the memory and read them frequently. If a read error occurs, it can be assumed that the victim has started a write operation near-by. This is true since the victim's read can't generate enough noise to cause failure in adversary's read operation.

### 5) FAULT INJECTION USING READ OPERATION

Fig. 12(a) shows the supply noise generated by various read data patterns. It can be noted that the adversary can control generated noise magnitude by reading specific data patterns (stored before to launch an attack). Fig. 12(b) shows that the sense margin reduces with supply noise. However, ground bounce affects more compared to droop as it, i) reduces the discharge current; and, ii) reduces VGS of access transistor (RTransistor is higher) while voltage droop only reduces discharge current. It is evident from Fig. 12(b) that if the adversary can generate supply noise in a way that the victim incurs noise > 150mV, the victim will read '1' incorrectly. However, injecting read error to data '0' requires significantly higher noise. Therefore, both polarity read failure (DoS by a read failure) might not be possible as the required noise is too high.

### 6) CONSIDERATIONS FOR OTHER NVMS

The NVM fault injection attack leverages supply noise. Since all NVMs incur high supply noise due to their high write current and their write current/time for data '0' and data '1' are asymmetric, all NVMs are susceptible to a similar fault injection attack.

### 7) COUNTERMEASURES

Following techniques can prevent or alleviate the attack:
i) Sequential read/write access: This can be a naïve solution as non-pipelined access hurts system throughput. However, the adversary will not be able to create droop/bounce or sense data by launching parallel access.

ii) Architecture-level mitigation: Parallel operations of different processes can be initiated to addresses with highest possible RInt. This will alleviate the issue to some extent.
iii) Good quality power/ground grid: A good power/gnd grid reduces supply line parasitics which in turn reduces bounce. However, this cannot eliminate the issue completely.
iv) Power rail separation for each bank: Separation of supply and gnd rails between parallel accessed banks will prevent the propagation of supply noise. However, this will incur significant area-overhead and reduce the power rail capacitance (which is not desirable).
v) System clock slow down: Higher TClock gives more time to read/write at lower headroom voltage to fix latency failures.

### C. INFORMATION LEAKAGE ATTACK

In this section, we describe an information leakage attack to approximate the HW of the victim's data by an aggressor in a shared computing environment share. RRAM is taken as a test case.

### 1) OVERVIEW OF INFORMATION LEAKAGE ATTACK BY SUPPLY NOISE

Fig. 13 shows the concept of an information leakage attack by leveraging supply noise. Victim writes sensitive data pattern which creates data-dependent supply noise and propagates to the adversary memory space. Adversary reads a known data (i.e. known supply noise) which adds up to the propagated noise and creates a read failure. From the read failure, the adversary can detect the amount of sensitive noise from the victim and back-calculate the HW of the victim's sensitive data. For example, if the victim writes a data that generates >150mV droop in the adversary's memory space, the adversary can conclude that the victim's write data HW is > 66.77% (with some assumptions).

### 2) ATTACK MODEL AND ASSUMPTION

It has been shown that an adversary can leverage high and asymmetric write current and long latency of NVMs (which causes supply voltage droop and ground bounce) to approximate HW of victim's write data [56]. An adversary can keep reading data in his memory space. If a read failure is incurred, the adversary can approximate the noise generated by the victim's write data which can further reveal a range of HW of the data. For this analysis, an RRAM-based (i.e., 1T1R) Last Level Cache (LLC) has been considered as a test case.

It has been assumed [56] that: i) NVM LLC is being shared by two users (i.e., an adversary and a victim); ii) bank level parallel read/write operation is performed to increase the throughput; iii) adversary has the knowledge of the amount of droop/bounce that can be generated by a read/write data pattern; iv) adversary also knows how the generated droop/ bounce propagates (decays with distance) and how it affects the victim's write/read operation; v) the adversary is an expert in computer architecture and can exploit knobs e.g., accessing specific data pattern in pre-

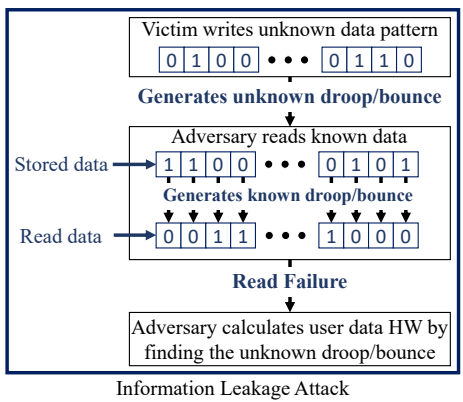

Information Leakage Attack

**FIGURE 13.** Information leakage attack using supply noise.

defined physical locations to prevent their replacement by policies e.g., Least Recently Used (LRU).

### 3) DATA INFORMATION APPROXIMATION

RRAM write waveforms can be divided into two regions with respect to time. The first region represents the old data state and the last region represents the new data state of the memory cells. Write current in the last region is approximately two values, namely, for data '1' (low current) and data '0' (very high current). The adversary can leverage these observations to approximate the HW of the victim's write data. Therefore, the adversary's WOI will be the last region once the write initiation is detected. The adversary can focus on the read/write failure characteristics in his memory space near-by to victim's write operation (after a write initiation detection as mentioned in Section III.2) and guess HW of victim's write data with some hypothesis. This information can be used in other attack models (e.g. SCA) to reduce the search space significantly, and thereby, improving the attack.

### 4) CONSIDERATIONS FOR OTHER NVMS

It has been noted that information leakage attacks leverage supply noise. Similar to RRAM, all NVMs incur high supply noise due to their high write current and their write current/time for data '0' and data '1' are asymmetric. Therefore, all NVMs are susceptible to similar information leakage attacks.

### 5) COUNTERMEASURES

The countermeasures against fault injection (discussed in Section III.2) by leveraging supply noise are also applicable for information leakage attack prevention by leveraging supply noise.

### D. ROW HAMMER ATTACK

In this subsection, NVM susceptibility to Row Hammer (RH) attack is discussed taking STTRAM as a test case. However, other NVMs also show similar susceptibility.

### 1) OVERVIEW OF RH ATTACK USING SUPPLY NOISE

RH on traditional memories e.g., Dynamic RAM (DRAM) [71] have revealed that it is possible to corrupt the data in nearby addresses by repeatedly reading from the same address. The authors have demonstrated this attack on Intel and AMD systems using a malicious program that generates many DRAM accesses. However, emerging NVM can also be vulnerable to RH attack. Such attacks have also been investigated on STTRAM [57] as a test case.

ATTACK MODEL AND ASSUMPTION:
It has been assumed [57] that STTRAM is designed similar to conventional embedded memories such as, SRAM and eDRAM which leads to high parasitic capacitance and resistance. The adversary can keep writing to particular addresses in his memory space. This results in high ground bounce due to high write current being dumped to the ground rail. This bounce will propagate to the word-line/source-line/bit-line drivers of the neighboring bits. If the bounce propagates to word-lines drivers, the unselected bits that share the same bit-line/source-line drives will partially turn the access transistor ON and a disturb current will pass through them. These bitcells will experience retention failure and read disturb. Furthermore, if the bounce propagates to source-line/bit-line drivers, the bitcells will experience lower voltage headroom. Therefore, read/write operations may fail.

EXPLOITING WRITE OPERATION:
For the RH attack, one particular address is written multiple times. This generates a ground voltage bounce (as mentioned earlier) that propagates to the peripherals such as, word-line, bit-line, and source-line drivers. These can cause the following issues:

i) Bounce propagates to word-line and source-line or bit-line drivers (causes retention failure and read disturb): The nearest unselected bits (in case of writing $0\square1$) whose source-line (or bit-line in case of writing $1\square0$) drivers share the same supply rails as bit-line or source-line drivers of the selected cells, have zero VGS at the corresponding access transistor since word-line and source-line (or bit-line) bounce together. However, the bounce propagates with a delay to the farther word-line drivers (due to different path delay) which results in a phase shift between bounce of word-line and source-line. Therefore, those access transistors will experience a brief period when the access transistors will weakly turn ON i.e., the VGS will be greater than 0V. This will introduce disturb current through those unselected cells, and they will eventually be written to either '0' or '1' (depending on the direction of disturb current) if the disturb current flows for a duration longer than the reduced retention time. Furthermore, if the attack takes place at elevated chip temperatures, the threshold voltage of the access transistor and the retention time of MTJ will be lowered, and the current through the MTJ will be higher leading to faster corruption of the bits and more effective attack. Therefore, by writing a particular address repeatedly, a massive number of unselected bits whose bit-line or source-line drivers share the same supply rails as bit-line or source-line drivers of the selected cells can be written/flipped.

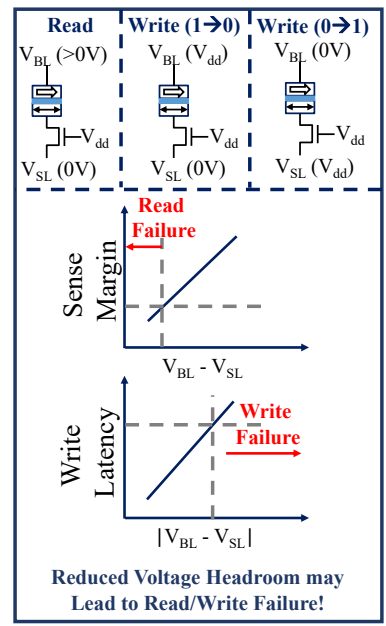

**FIGURE 14.** Reduced voltage headroom decreases sense margin and increases write latency and may lead to read/write failure.

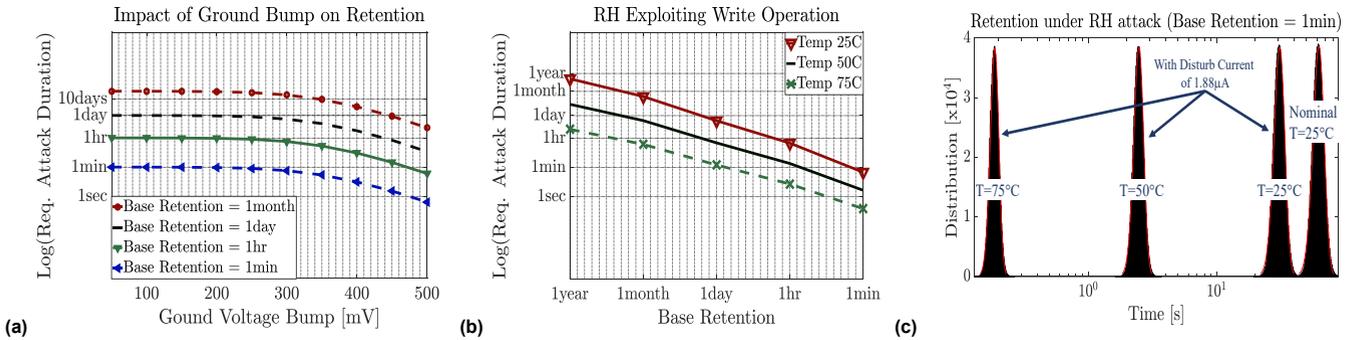

**FIGURE 15.** (a) Impact of ground voltage bounce on retention time of unselected bits (base retention = 1month); (b) impact of RH attack on STTRAM write operation for different base retention time; and, (c) retention distribution under RH attack for base retention = 1min. A 1-million-point Monte-Carlo analysis is conducted with 3σ of 2% of MTJ thermal stability factor, Δ0 and with a mean of Δ0=24.85 (corresponding retention time ~1min).

It should be noted that the disturb current lowers the thermal barrier. Therefore, if the partially selected bits in other independent banks are read, the probability of a read disturb of these bits increases. Note that process variation can amplify these issues further since the weak unselected bits (with lower thermal stability) and low access transistor threshold can easily get corrupted.

ii) Bounce propagates to source-line only (read failure): Let's assume that adversary is writing in a bank and generating ground bounce and the victim is reading data from another independent bank. Therefore, the bitcells that are being read by the victim will have a zero source-line and non-zero word-line and bit-line (Fig. 14). If the bounce generated by the adversary reaches the bitcells that are being read, the read operation will incur a lower sense margin due to lower voltage headroom (source-line voltage bounces) (Fig. 14). Therefore, read failures may occur if the sense margin degrades significantly.

iii) Bounce propagates to source-line or bit-line (write failure): Let's assume that adversary is writing in a bank and generating ground bounce and the victim is writing in another independent bank. Therefore, the bitcells that are being written by the victim will have a zero source-line (for 1→0 writing) or bitline (for 0→1 writing) (Fig. 14). If the ground bounce generated by the adversary reaches those bitcells, the write operation will incur longer write latency due to lower voltage headroom (source-line or bitline voltage bounces) (Fig. 14). Therefore, write failures may occur if the increased write latency is greater than the design target.

RETENTION, READ AND WRITE FAILURE:

Retention Failure: Fig. 15(a) shows that as the ground bounce seen by the bitcell increases, the retention time of the cell reduces. It has been shown that a higher temperature can reduce the retention time further (Fig. 15(b)). The RH attack can flip the bits in ~30secs at T=25°C if the base retention is 1 min which can be reduced to 2.5secs and 0.2secs at T=50°C and T=75°C, respectively. Fig. 15(c) shows the weaker bits under process variation are more vulnerable to the attack since their retention reduces to ~19secs, ~1.7secs, and ~0.1secs at T=25°C, T=50°C and T=75°C respectively.

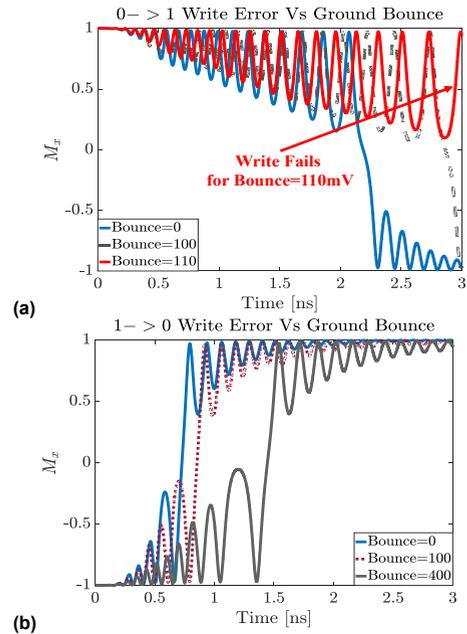

**FIGURE 16.** Write time for (a) 0→1; and, (b) 1→0 increases as the bitcell being written experience higher ground bounce.

**Read Disturb:** A small read current is passed through the bitcells during a read operation. Higher read current gives better sense margin but increases read disturb probability. Therefore, the read current is selected in a way that does not flip the bit as well as yields a good sense margin. However, disturb current due to ground bounce lowers the thermal barrier of the bitcell. If the bitcell is read at the lower thermal barrier, switching probability during a read operation (read disturb) increases [21]. Furthermore, higher temperature further increases the switching probability.

**Read Failure:** Read failure may occur if the bitcell being read experience ground bounce (generated by parallel access in another independent bank) in its source-line. respectively. It is shown that sense margin for data 1 reduces whereas sense margin for data 0 stays relatively constant as the ground bounce experienced by a bitcell during a read operation increases. A lower sense margin can lead to an incorrect read.

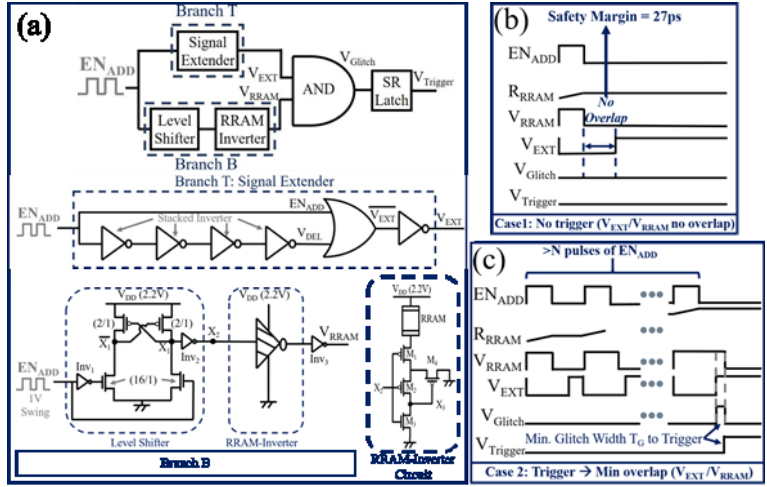

**FIGURE 17.** (a) One flavor of the dealy-based NVM Trojan trigger; waveform (b) before; and, (c) after N times of hammering (generates Trojan signal).

**Write Failure:** Write failure may occur if the bitcell being written experience ground bounce (generated by parallel access in another independent bank) in its source-line (for writing 1→0) or bit-line (for writing 0→1). Fig. 16(a) and (b) represent the impact of ground bounce on 0→1 and 1→0 writing respectively. Writing 0→1 fails if the bitcell experience 110mV of ground bounce as the magnetic orientation ($M_x$) is not reaching -1 (anti-parallel state). However, 1→0 write failure might not be possible as even with 400mV of ground bounce the magnetic orientation ($Mx$) successfully reaches 1 (parallel state). Therefore, 1→0 write failure requires high ground bounce which might be possible to generate by a parallel write operation.

At first glance, the RH attack on STTRAM might not seem severe compared to DRAM. However, in contrast to the RH attack on DRAM that only causes data corruption by retention failure, the RH attack on STTRAM can cause data corruption (retention failure, read disturb) and fault injection (read/write failure).

CONSIDERATION FOR OTHER NVMS:

The NVM RH attack leverages supply noise. Although STTRAM is taken as a test case, we note that other NVMs also incur high supply noise due to their high write current. Therefore, all NVMs will incur read and write failure due to high supply noise propagated from a parallel write operation. PCM and RRAM are susceptible to resistance drift. Therefore, the RH attack can change their resistance by few ohms if a current pass through the cell during every hammering. Eventually, the cell content may get corrupted. However, the impact of supply noise on the retention time of PCM, RRAM and FeRAM is not investigated. This could be a topic of future research.

MITIGATION TECHNIQUES:
The countermeasures mentioned to prevent fault injection by leveraging supply noise are also applicable to the RH attack by leveraging supply noise. Additionally, write operations can be stalled to facilitate recovery of lost retention to mitigate the susceptibility of STTRAM to RH attack. The average disturb current reduces by 80% by stalling write operation by one cycle after every four consecutive writes which in turn increases the attack duration to 3.2secs (1.30X improvement) and 0.3secs (1.57X improvement) for 1 min of base retention at T=50°C and T=75°C respectively.

### 2) RH BASED DOS ATTACK

NVMs such as, PCM and RRAM suffers from endurance issues. The oxide layer in the bitcell of PCM/RRAM breaks down after a specific number of write cycles. The endurance cycle of RRAM and PCM are significantly lower than other NVMs. Therefore, an adversary can keep writing new data to cache memory addresses and lower their endurance. This will lead to a DoS attack when the life cycles of the bitcells get expired.

COUNTERMEASURES:

Wear-leveling techniques can be implemented to prevent DoS attack on NVM-based LLC. For example, logical to physical address can be mapped dynamically based on the number of cycles they are already written. A number of wear-leveling techniques are proposed for NVM-based main memory (details in Section VI.1) which can be extended to NVM-based LLC.

## IV. NVM-ENABLED TROJAN ATTACKS

In this section, we present various Trojan attacks by leveraging NVM characteristics.

### A. Emerging NVM-based Trojan Triggers (ENTT)

In [72], a delay (Fig. 17(a)) and voltage-based NVM Trojan trigger are proposed by exploiting the RRAM resistance drift under pulsing current (Fig. 18). The basic idea is to hammer an RRAM cell which increases its resistance by a few ohms per hammer (Fig. 18). This increment of resistance can be converted to increase in delay (implementing the RRAM in an inverter) or decrement of a node voltage (by implementing the RRAM in a voltage divider (Fig. 17(a))). Before the hammering, the delay is below a threshold value and trigger output remains zero (Fig. 17(b)). Once the delay (voltage) is above (below) a predefined threshold (Fig. 17(c)), the circuit will generate a pulse that can be captured by an SR latch (Fig. 17(a)). This serves as the Trojan trigger.

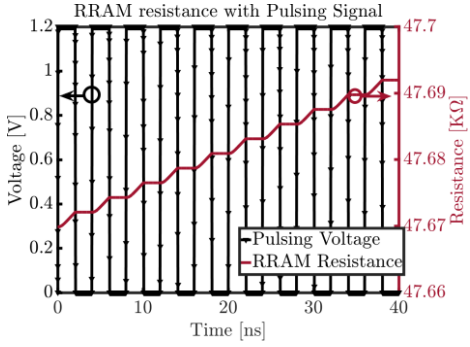

**FIGURE 18.** RRAM resistance drift with pulsing current.

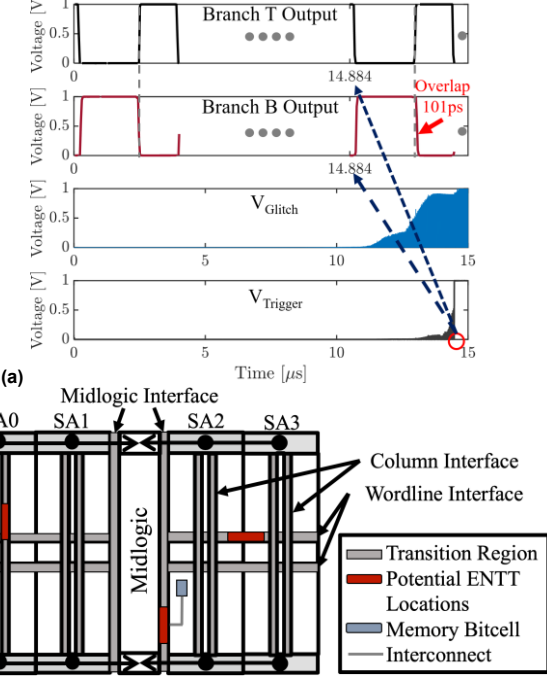

**FIGURE 19.** (a) VTrigger is asserted after 2500-3000 times of hammering; (b) placement of ENTT within RRAM memory array.

Simulation results indicate that these triggers can be activated by accessing a pre-selected address 2500-3000 times (varies with trigger designs). The proposed trigger evades the test phase since it requires a large number of hammerings (Fig. 19 (a)).

Due to RRAM's non-volatility, the hammering could be rare and therefore, can evade system-level techniques that can classify hammering as a potential security threat. The work also describes that the trigger circuit can be implemented in the peripheral of the memory array (Fig. 19(b)). Sacrificial bitcells are used at the interface of array and column areas and, array and wordline areas for a smooth transition to logic and to maintain high yield. These non-functional sacrificial bits can be repurposed to hide the Trojan trigger (by the designer or the fabrication house). The additional logic e.g., inverter chains and/or comparators can be hidden in the filler areas of the non-memory logic (e.g., address pre-decoding and pipelining units), also called midlogic and connected to the RRAM. Floating metals are abundant in the address generation logic and can be reused to route the trigger signal without causing any area-overhead. This makes it difficult to detect the Trojan via optical inspection.

The Trojan trigger consumes dynamic energy only during the hammering of a predefined address. The static power consumption is significantly less. Therefore, analyzing the power spectrum and comparing it with a golden chip may not be able to detect the Trojan. This makes such NVM Trojan extremely dangerous.

In [73], a capacitor-based Trojan trigger (Fig. 20) for NVM is proposed which is small, sneaky and stealthy. Adversary hammers a pre-defined memory address with a predefined data pattern. Every hammering increases the charge stored in a capacitor. If the capacitor gets charged more than a threshold value, it generates a signal which can be considered as the Trojan trigger signal. The advantage of this capacitor-based Trojan trigger is it requires a large number of hammering and therefore, can evade the detection during the testing phase. An optical inspection may not work since the Trojan circuit is small and sneaky. The circuit also consumes low static power which makes it difficult to detect via power spectrum comparison with a golden chip. However, the limitation of such a capacitor-based Trojan is the hammering requires to be fairly continuous. If hammering is stopped for a sufficiently long period of time, the capacitor may get completely discharged.

## B. Trojan Payloads for NVM

Once triggered, the Trojans proposed for NVMs can launch the following attacks:
(i) Information leakage (Fig. 21(a) [73]): It is assumed that the victim and adversary have control over WL[0] and WL[1], respectively. The WLs share the same bit-line (BL[0]) and source-line (SL[0]) and are coupled through a Trojan transistor (switch). If the switch is activated by a Trojan trigger, the data will be copied to WL[1] whenever the victim writes to WL[0]. The adversary can read WL[1] to leak the victim's write data. (ii) Fault injection: The Trojan can target memory addresses to prevent writing one particular data polarity (either $0\rightarrow1$ or $1\rightarrow0$). In Fig. 21(b) [73], we note that $0\rightarrow1$ fails since the headroom voltage between bit-line and source-line is not sufficient to write the cell [55-56]. However, writing $1\rightarrow0$ is successful. As mentioned before, such fault injections can leak system assets such as, cryptographic keys.

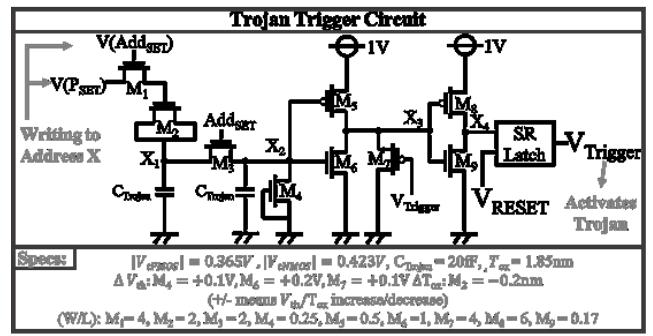

**FIGURE 20.** Capacitor-based Trojan trigger circuit.

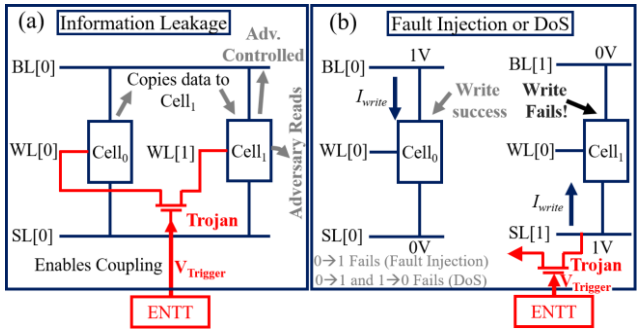

(iii) DoS: If Trojan targets both write polarities ($1\rightarrow0$ and $0\rightarrow1$), the victim will not be able to write anything to the memory. This results in a DoS attack.

### C. Considerations for Other NVMs

In this work, an RRAM-based NVM Trojan is summarized which can evade detection during the testing phase and system-level countermeasures. However, note that similar Trojan can be designed using PCM memory since they also exhibit resistance drift. In general, STTRAM and MRAM cells exhibit two resistance states and not suitable to design NVM Trojan. FeRAM resistance can change in a certain small range since the ferro-electric capacitor charge can vary. However, such FeRAM-based NVM Trojan will trigger with significantly smaller number of hammering which makes it suitable to be detected using testing techniques.

### D. Countermeasures

The following countermeasures can prevent the attack:
(i) Small validated ECC: A carefully validated and optically inspected ECC (free of Trojan) can be used to store the ECC for each memory word. If the Trojan performs fault injections/DoS, the ECC will detect it.
(ii) Analysis of memory images: Memory Trojans are visually tedious to identify due to replication of a large number of memory instances. Machine learning can be applied to analyze the memory bank images to identify anomalies. This approach will worsen the test/validation time.
(iii) Temperature/voltage modulation to screen Trojan: Higher operating voltage will accelerate the drift of the RRAM's resistance in the trigger circuit. Therefore, the Trojan could be triggered quickly and can be detected. Similarly, higher temperatures will lower the HRS of the RRAM and aid in detection.

## V. SECURITY ISSUES OF NVM CACHE

In this section, the security issues related to NVM-based cache and countermeasures are discussed.

### A. Tampering and DoS

In this subsection, NVM susceptibility to tampering is discussed taking STTRAM as a test case. However, other NVMs also show similar susceptibility.

#### 1) TAMPERING AND DOS ON STTRAM

STTRAM is susceptible to contactless tampering efforts, e.g. by subjecting it to a strong external magnetic field and/or thermal field, an adversary can corrupt stored contents [58]. The PL of MTJ in STTRAM is robust. However, the FL of MTJ could be toggled through both spin-polarized current as well as a magnetic field. The FL is susceptible to both the magnitude and polarity of the external magnetic field it is subjected to. The motivation of tampering for an adversary is to corrupt the data or steal information. This could prevent STTRAMs application to a wide range of mobile devices. Note that although the work takes STTRAM as a motivational case study, other forms of magnetic memories such as, MRAM is also expected to experience a similar issue, and hence vulnerable to tampering attacks.

#### 2) ATTACK MODEL

The attacks on STTRAM could be launched either through static (DC) magnetic field or alternating (AC) magnetic field. The DC attack is less detrimental as it can only create unipolar failures. For example, a magnetic field will cause failures only for the bits whose FL layer orientation is opposite to the applied field. However, the AC field could cause more damage as it will affect both storage polarities. Due to ease of AC field generation using a low-cost electromagnet this type of attack is highly likely.

The attack could be launched either during retention mode or read/write (functional) mode. Note that read current is unipolar irrespective of the storage polarity whereas write current polarity is data-dependent. The impact of attack during functional mode (especially read) could be more detrimental than retention due to two factors: (a) presence of disturb current; and, (b) higher frequency of reads compared to writes. Both storage polarities will be affected under AC attack. During a write operation, the AC field will either assist if the current polarity matches with the magnetic field or suppress the attack if the current polarity is opposite to the applied field. In all of the above scenarios, the attack could either manifest as a hard failure (i.e., flipping of the bitcell content) or soft failure (i.e., delay in write or degraded sense margin). The soft failures could be mitigated by slowing down the read/write operation but the hard failures need to be avoided or corrected through error correction.

The frequency of the magnetic field is important in the context of failures in functional mode. If the AC field frequency is faster than the write time then it can affect writing both data polarities. Similarly, it can also affect both storage polarities during a read operation. If the frequency of the AC field is slow then the impact will be less harmful.

MTJ FL magnetic orientation could be flipped in retention mode also. The flip time reduces with the increase of the strength of an external magnetic field. Higher frequency AC field can cause more damage even with smaller amplitude than lower frequency AC field and higher amplitude.

For the DC field, the bits can fail easily when the current polarity and magnetic field are in the same direction (assistive). The flip time is higher when the current and the magnetic field are in the opposite direction (suppressive). The similar conclusion also holds true for AC field.

The stability of MTJ FL is a function of its volume. Therefore, it is possible to enhance the robustness of the MTJ against tampering by increasing the size. It is shown in [58] that the bitcell is able to withstand a weak magnetic attack with a higher volume of MTJ FL. However, it fails to provide protection against a stronger attack (>400Oe). A higher volume of FL of MTJ can protect the cell against an attack of lower frequency. High-frequency attack can cause failure regardless of MTJ FL volume. Therefore, retention mode can be considered more robust to attack compared to functional modes.

In [74], the authors consider two types of magnetic attack on STTRAM LLC. In the first case, the strength of the attack ramps up gradually and in the second case strength of the attack ramps suddenly. The gradual ramping attack is more practical when a human entity is involved in the attack process and a permanent magnet or electromagnet is brought closer to the memory manually. The adversary can launch DoS attack by bringing a permanent magnet close to mobile devices such as IoTs and cell phones. The sudden attack applies to scenarios where the adversary has physical access to the memory and has precise control over the magnetic field strength and proximity from the chip. An insider in a computer facility can launch DOS attack by physically accessing the memory. In sudden attack, the functional bits can fail immediately if the field strength is beyond the threshold value.

### 3) CONSIDERATIONS FOR OTHER NVMS

Tampering and DoS attack exploits external magnetic and temperature. All NVMs are susceptible to the external thermal field while spintronic memories are susceptible to both external magnetic and thermal fields. Furthermore, FeRAM is susceptible to an external electric field. Therefore, the conclusion drawn in this section holds true for other NVMs as well.

### 4) TAMPERING AND DOS ON STTRAM

MTJ that flips faster compared to the usual memory bits can be implemented to detect the attack [58]. Once detected, the following techniques can be implemented to mitigate the attack.

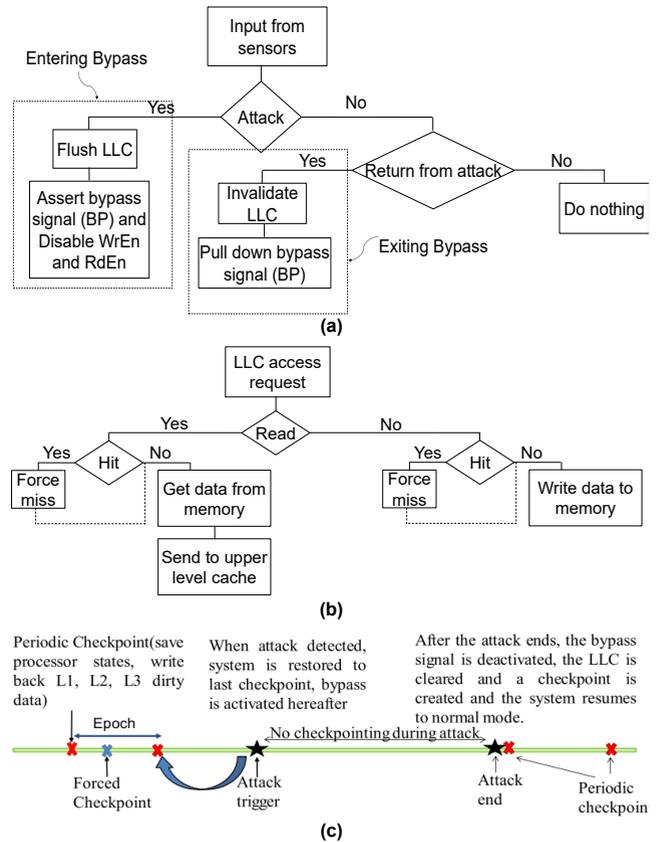

**FIGURE 22.** (a) Control flow to activate/deactivate bypassing; (b) processing of read, write requests during bypassing; and, (c) cache bypass architecture with checkpointing.

(i) Array Sleep: The work [58] proposes to put the memory in retention mode since retention mode is more susceptible to attack. However, a strong enough attack can corrupt all the stored information even in the idle mode.

(ii) ECC: Fixed and variable length ECC can be implemented to correct the corrupted bits. However, this method will fail if more bits are corrupted compared to the strength of ECC.

(iii) Stalling: CPU can be stalled and wait till the attack is over. If the cache implements a write-back policy, then the dirty data is written back to the main memory to save the system state on detection of the attack (for gradually ramping attack) and the CPU is stalled. After the attack is over, the entire LLC is invalidated and the computation starts from the last saved state. The processor's register contents will remain intact and the computation can resume from the state it was halted. This technique is better than shutting down the entire system because the processor states remain intact and the computation can instantly start after the attack is over. For the user, the machine will appear to be stuck during the attack however, the user is not required to reboot the system. Although simple, this technique will not work for a sudden attack since the dirty data will be corrupted. For such scenarios, the processor has to be restarted after the attack and the applications can restore the states if application-level checkpointing [75-76] is implemented. These methods prevent DoS attack successfully as the system does not consume corrupted data. However, both approaches disable computations during an attack and result in power loss. The attacker can also exploit these features to drain the battery of the system.

(iv) Cache Bypass [74]: Cache bypassing enhances the user experience as the computation continues with affordable IPC degradation. Fig. 22(a) shows the necessary steps needed to prepare for bypassing, continue bypassing and exit bypassing. If the sensors indicate a weak attack the LLC is flushed by copying the dirty data and a bypass signal (BP) is asserted. In absence of an attack, if the bypass signal is still asserted (indicating the end of attack), the entire LLC is invalidated and the BP signal is de-asserted. Otherwise, no extra steps are needed. The entire bypass operation has two steps:

- Preparing for bypassing (Fig. 22(a)): If the sensors indicate an attack, the dirty data in LLC is copied to the main memory by asserting the FLUSH signal [77] in the cache controller to ensure correctness. Note that this is possible since the sensors can sense the attack before bits start failing [58]. The FLUSH signal writes back the dirty blocks and invalidates all the cache lines after the write-back. The BP is asserted to indicate the cache controller to bypass the subsequent requests to the main memory. Note that if LLC employs write-through policy then this step is not necessary as the copy of data is immediately written back to the main memory;

- Bypass mode (Fig. 22(b)): There are four scenarios when the data can leave or enter the LLC namely, read-hit, read-miss, write-hit and write-miss. The read hits are forcibly converted to read misses so that the data is read from the main memory instead of the cache. Read misses are served normally by sending the data from the main memory. Write hits are also forcibly converted to write misses and the data is written only to main memory. In the case of write misses the main memory is updated with the new data. During the attack, LLC data should not be used for computation or stored anywhere (upper-level caches, main memory). Note that new data may be read (written) from (to) the LLC and discarded during bypassing which results in energy overhead. In order to save dynamic energy, the LLC is prevented from performing read or write operations. This can be done by ANDing $\overline{(BP)}$ with WrEn (write enable) and RdEn (read enable) signals which is generated by the cache controller.

(v) Checkpointing [74]: System-level checkpointing can be leveraged to mitigate the sudden attacks. The CPU register values and PC are saved in hard drive. Additionally, LLC dirty blocks are stored in the main memory. When an attack is sensed the system is restored to the last saved checkpoint and the bypass signal is asserted (Fig. 22(c)). The system continues to perform with the LLC bypass and the checkpointing is disabled to avoid write back of stale LLC data. After the attack ends, the bypass signal is de-asserted, the LLC is invalidated and a checkpoint is created. The system continues to perform normally with checkpointing resumed.

### B. DoS using Supply Noise

In this subsection, NVM susceptibility to DoS attack by leveraging supply noise is discussed taking RRAM as a test case. However, other NVMs also exhibit similar susceptibility.

#### 1) BOTH POLARITY WRITE/READ FAILURE

In Section III.2, it has been discussed that the adversary can generate supply noise by writing in his memory space and cause write failure. If the failure is for both polarity of data, it leads to DoS attack. Both-polarity-read failure using supply noise requires very high noise which might not be possible to generate. However, the adversary can use other voltage sources to inject supply noise and launch DoS attacks on the read operation.

#### 2) CONSIDERATIONS FOR OTHER NVMS

DoS attack leverages supply noise and all NVMs incur high supply noise due to their high write current. Therefore, all NVMs will incur both polarity read and write failure due to high supply noise propagated from a parallel write operation.

#### 3) MITIGATION TECHNIQUE

The countermeasures mentioned to prevent fault injection by leveraging supply noise are also applicable to DoS attack by leveraging supply noise.

## VI. SECURITY AND PRIVACY ANALYSIS OF NVM MAIN MEMORY AND STORAGE

In this section, security and privacy analysis of NVM-based main memory and storage and countermeasures are presented.

### A. Analysis of NVM Main Memory

In prior works, PCM-based main memory has been investigated for their security and privacy issues. A number of secure designs are also proposed. This section summarizes these techniques.

#### 1) iNVMM [78]

An adversary with physical access to the system which implements NVM as main memory, can extract sensitive information long after the system is powered down due to the persistent nature of NVM data. Therefore, a unique data privacy protection technique called i-NVMM is proposed where the main memory is encrypted incrementally. This means that different data is encrypted at different times depending on whether the data is predicted to be useful to the processor. This is done by predicting the 'inert' pages of the main memory by scanning them periodically and identifying the ones that haven't been used for a long time. Once identified, these 'inert' pages will be encrypted by a memory-side encryption engine. Since the encryption is done periodically and does not wait for a power event, the attack window when the adversary can steal information is significantly less. Furthermore, since the encryption is done on the memory side, it does not rely on specific processor architecture and incurs zero system performance overhead. Simulation result indicates that i-NVMM encrypted across SPEC2006 benchmarks results in 3.7% execution overhead and with negligible impact on NVM write endurance.

## 2) IMPROVING PRIVACY AND LIFETIME [79]

Although [78] claims that data encryption impacts write endurance negligibly, encrypted data is randomized due to its diffusion characteristics [79] which negates some of the prior proposed wear-leveling techniques such as, redundant bit-write [80] and partial write [81]. Therefore, two methods are proposed to reduce the impact of encryption on the wear-leveling techniques [79]: i) extension to encryption scheme which revives partial writes; ii) implementing age counter by leveraging encryption counter and dynamically adjust error protection strengths. For the extension of the encryption scheme, the technique [79] adds multiple block-level counters in addition to the cache line counter for each cache line. After encryption when a write-back is done, only the dirty blocks are written and their counters are incremented. Simulation result indicates that the new encryption scheme can improve memory cell lifetime by ~2X at 1.6% area overhead.

## 3) DEUCE [82]

A Dual Counter Encryption technique, namely DEUCE [82] is proposed which leverages the fact that a typical write-back only changes a few words. Therefore, DEUCE encrypts only those changed words. This nullifies the necessity of handling the full-length encryption after every write back and thereby, increases performance. Furthermore, it also leads to an improvement of the overall lifetime of memory cells since not all cells are written after every encryption. Simulation results indicate that this technique reduces the number of bits that are modified per writeback from 50% to 24%, and improves performance and lifetime by 27% and 2X, respectively.

## 4) EFFICIENT CHECKPOINTING OF LOOP-BASED CODES [83]

The technique in [83] investigated the impact of different checkpointing schemes on loop-based codes on NVM main memory. The results indicate that logging applied to a title loop increases the number of the write operations to NVM main memory. This, in turn, reduces the write endurance. Therefore, [83] proposes a recompute-based technique that only logs sufficient states to enable correct re-computation. If a failure is observed, this approach can recover to a consistent state by going back to failed computation. This nullifies the necessity of continuous checkpointing or logging and thereby, improves the write endurance and reduces execution time. Simulation result indicates that this approach reduces execution overhead from 8% (with logging)/207% (with checkpointing) to 5% and reduces write to NVM main memory from 111% (with logging)/330% (with checkpointing) to 7% for tiled matrix multiplication.

## 5) ENHANCING LIFETIME AND SECURITY WITH START-GAP WEAR LEVELING [84]

The achievable lifetime of PCM-based main memory can get reduced by 20X due to the non-uniformity of a write operation to different cells. Although this can be mitigated by wear-leveling, the solution requires large storage tables and indirection which incur significant overheads. It has been noted [84] that the storage table can be eliminated if an algebraic mapping is used between logical and physical memory. A technique called Start-Gap is proposed which uses two registers namely, Start and Gap. After a specific number of writes to main memory, Start-Gap moves one line to a neighboring location. Logical to physical mapping is done by a simple arithmetic operation of Gap and Start registers with the logical address. Simulation result indicates that this technique can improve the achievable lifetime of PCM cell from 5% of the maximum possible lifetime to 53%. Furthermore, Start-Gap incurs a total storage overhead of less than eight bytes and limits extra write caused by wear-leveling to <1%. This technique can also prevent failure caused by repeated writes (to the same line) based attacks.

## 6) ONLINE ATTACK DETECTOR (OAD) [85]

The technique [86] suggested that the Start-Gap wear-leveling technique [84] is vulnerable to Birthday Paradox Attacks. This is a type of cryptographic attack that exploits the probabilistic model to reduce the complexity of finding a collision for a hash function. To prevent such attacks, [85] proposes a novel OAD circuit which can adjust the wear-leveling algorithm based on memory reference stream properties. This is done by introducing an attack detection notion by identifying memory access patterns that are malicious. OAD incurs hardware overhead of few tens of bytes, however, can protect PCM-based main memory from a large family of attacks.

### B. NVM-based Storage in IoTs

In this section, the issues related to replacing eFlash in IoTs with STTRAM are discussed. A novel memory architecture is also described which can mitigate the issues.

## 1) TAMPERING AND DOS WITH EXTERNAL MAGNETIC AND TEMPERATURE [58]

As mentioned earlier, STTRAM and MRAM are susceptible to contactless tampering efforts, e.g. by subjecting it to a strong external magnetic field and/or thermal field [58]. Furthermore, all NVMs are susceptible to thermal attack. Therefore, contactless tampering is a security threat for NVMs. In Section V.1, tampering NVM-based LLC is explained by taking STTRAM as a test case. Although a similar conclusion holds true for NVM-based storage in IoTs, the mitigation techniques are different for them. For example, tampering LLC during the active mode of operation is critical than tampering in power-down mode. This is true since the LLC is always invalidated at power on. However, when STTRAM is used to store the application program (such as an eFlash replacement), attacks during both active and power-down mode become critical. This is true because the integrity of the program memory data needs to be maintained throughout the power cycles for the correct functionality of the IoT after each power on. In the LLC application, the main memory acts as the backup since it contains a copy of the LLC data from which the corrupted LLC bits can be recovered. However, this is not true for program memory due to a lack of backup data. Existing memory resiliency technologies like error-correcting code (ECC) as a standalone solution are not sufficient to recover the corrupted bits as they can only recover random bit errors, radiation-induced errors, or dynamic variability errors.

### 2) NON-INVASIVE MAGNETIC ATTACK ON IOTS WITH STTRAM [87]

In [87], the challenges related to replacing eFlash in IoTs with STTRAM is investigated. The work only considered a non-invasive magnetic field attack and assumed that only one IoT in a homogenous network is under attack at a time for simplicity. This situation is likely when multiple IoTs are distributed in a building or critical infrastructure such as a bridge, to collect the required information. Fig. 23 explains the attack sensing and recovery process. During the normal operation of the IoT, if an adversary tries to attack the STTRAM with the intention to scramble the stored firmware, the attack sensors and the Integrity Checker is able to sense the attack ahead of time. The STTRAM Integrity Checker detects the scrambled sensor arrays and sends the HALT interrupt as an attack signal to the microcontroller. If the adversary tries to launch the attack when the IoT is powered off, the passive sensors are able to detect the attack due to the failure of sensor bits. When the IoT is powered up after the attack, the boot ROM triggers the STTRAM Integrity Checker and the STTRAM will fail the integrity check due to the modified sensor arrays from the previous attack. The boot ROM then sets the IoT working mode to Support Request and starts executing the recovery request code from the EPROM.

With support requests, the IoT under attack requests the firmware. An IoT which is not affected by the attack sends a valid firmware over the UART connection. When the firmware transmission is complete, the sensor arrays are reset to its original configuration of alternating '0' and '1'. The recovery assist routine on the assisting IoT after transmitting the entire firmware from the STTRAM reboots the IoT in normal operation mode.

Since the entire recovery procedure is a critical operation, it needs to be safeguarded against any potential data leakage and unauthorized access. A wired connection such as serial USB and Ethernet is preferable in this case compared to a wireless network interface like WiFi or Bluetooth.

### 3) MITIGATION AGAINST NON-INVASIVE MAGNETIC ATTACK [58, 87]

It has already been mentioned that ECC as a standalone countermeasure is not sufficient. However, the authors have noted that the proposed recovery mechanism incurs significant

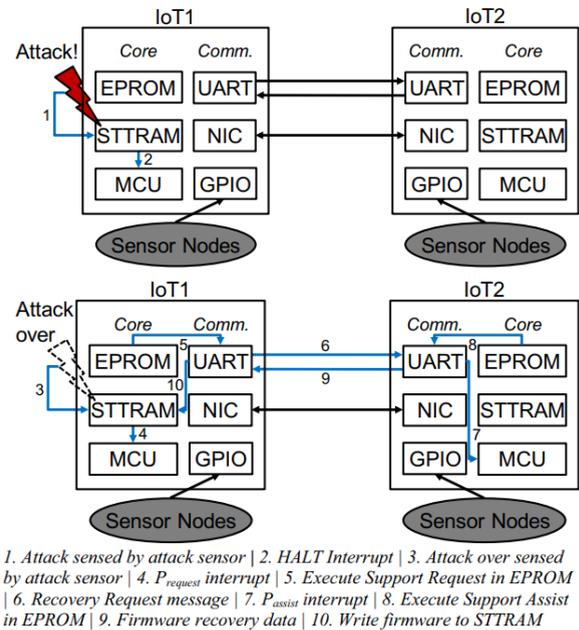

*1. Attack sensed by attack sensor | 2. HALT Interrupt | 3. Attack over sensed by attack sensor | 4. $P_{request}$ interrupt | 5. Execute Support Request in EPROM | 6. Recovery Request message | 7. $P_{assist}$ interrupt | 8. Execute Support Assist in EPROM | 9. Firmware recovery data | 10. Write firmware to STTRAM*

**FIGURE 23.** Attack sensing and recovery. The sequence of events is numbered and explained.

energy overhead especially considering the limited resources of an IoT. Therefore, the authors have proposed the following techniques to make the design robust and initiate the recovery routine only when the attack is strong enough which cannot be mitigated:

**i) Retention Enhancement:** STTRAM retention time can be incremented making the bits more resilient to failure in the presence of external magnetic fields. This, in turn, will reduce the number of times the recovery process needs to be invoked. The retention time of the STTRAM increases exponentially with the increase in the thermal stability. This can be achieved by increasing the volume of the MTJ free layer. However, this is limited by the saturation point of the thermal stability factor. Higher retention time is critical to enhancing the resilience of STTRAM against the magnetic field. However, higher retention time is associated with higher write energy. Therefore, a trade-off can be made between attack resilience and write energy.

**ii) ECC:** The application of ECC on STTRAM as a mechanism to further reduce the need for the expensive recovery process. Variable strength ECC [58] can lead to a reduction of bit error rate (BER) in STTRAM. BER can be reduced by appending correction or parity bits to the words stored. Noted that ECC cannot fix massive memory errors; however, it can fix random bit errors. ECC encoding will be only employed once during write and correction will be needed only after an attack event has been detected and subsided. The normal read operation will not require ECC and therefore will not experience any latency overhead due to ECC. When an attack event is sensed using the detection sensors, the memory will be read, corrected using ECC and written back. If the error cannot be fixed using ECC, then the corresponding memory chunk will be fetched from neighboring IoT, as per the recovery procedure. If all errors in the memory chunk are fixed using ECC, then it will not incur the recovery overhead.

## VII. THREATS ON COMPUTE CAPABLE NVMS

In this section, we present the state-of-the-art IMC design with NVMs, Next, we present their vulnerability and the attack models by which the IMC can be corrupted.

### 1) IN-MEMORY COMPUTATION USING NVMS

(i) Dynamic Computing In-Memory (DCIM) [88]: DCIM is a low-power dynamic computing in-memory technique that implements any logical function in the Sum-of-Product (SOP) form in RRAM crossbar arrays. The functions are executed in two steps: (a) AND'ed product computation and (b) OR'ing them. RRAMs in the crossbar array are pre-programmed to perform a particular function. Operands of an AND (OR) in a specified BL are in LRS and all other RRAMs are programmed to HRS. Initially, BLs are pre-charged to VDD. Once the enable is asserted, the BL voltage drops if any of its operands is '0'. Finally, BLs are pre-discharged to GND and BL will be charged if one of is inputs is logical '1'. Implementation of the XOR function using DCIM is shown in Fig. 24.

(ii) Floating Point Adder using IMC (FAME): FAME uses RRAM crossbar arrays to compute Floating Point (FP) operations in the memory. Instead of implementing functions in the AND-OR form, FAME implements functions in NAND-NAND and NOR-NOR forms. FAME carries out FP addition/subtraction in three steps: (a) In exponent subtraction, two FP numbers' exponents are subtracted from each other and the bigger exponent is detected based on the output's sign. Then, the fraction of the smaller number is shifted to the right by the difference of exponents; (b) In fraction addition, the shifted fraction and the bigger number's fraction based on their sign and type of operation are added/subtracted with/from each other; (c) In normalizing operation, the computed fraction from the previous stage is transformed into a FP presentation. FAME proposes an architecture that needs only left shifts in this step. FAME also proposes a new SA to enable shifting within memory arrays. This SA is capable of shifting both to the right and left based on the pre-programmed array connected to it.

(iii) In-memory Floating Point Computations for Autonomous Systems (FPCAS) [89] extends FAME to perform FP multiplication. The multiplication is done in three stages: (a) Exponents of the two FP numbers are added together, (b) the exponent is normalized into a FP presentation (a bias is added to exponents of FP numbers: FPex = ex + bias. When two fractions are added together, the bias should be subtracted (FPex1+FPex2 = ex1 + bias + ex2+bias, (c) fractions of the two FPs are multiplied together and the result is normalized.

(iv) SHA-3 Implementation In-Memory Computing (SHINE) [90]: SHINE is a high performance and area efficient hardware implementation of the Keccak function that forms the core of SHA-3 by exploiting RRAM-based IMC and implementing its Keccak function in a SOP form in the crossbar array architecture. Keccak consists of five steps that involve the application of one or more of the Boolean bitwise operators of XOR, SHIFT, AND, and INVERT. Each step is allotted dedicated arrays and register files to store intermediate hashed states and function simultaneously on different message blocks to ensure pipelined computation Fig. 25.

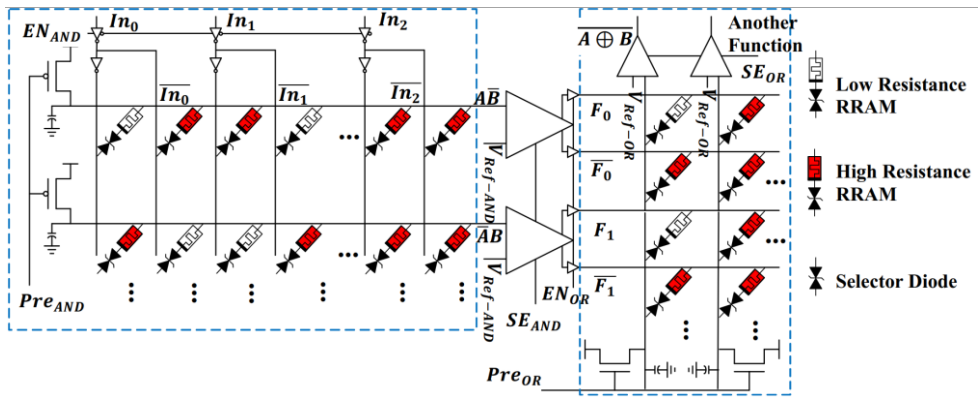

**FIGURE 24.** XOR implementation using DCIM [88].

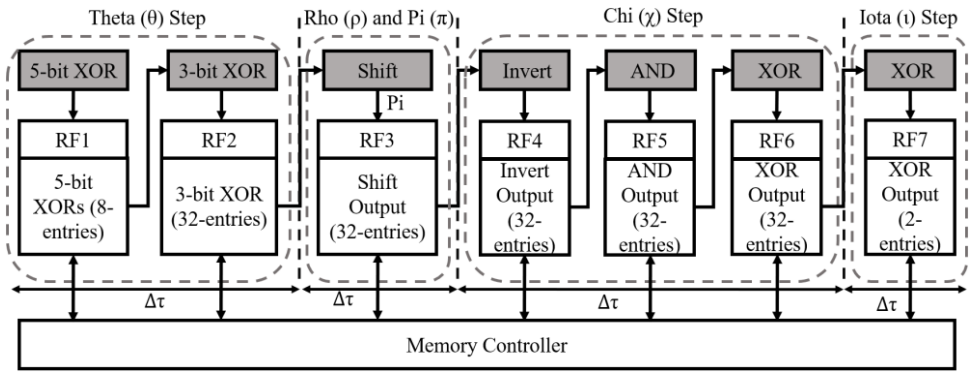

**FIGURE 25.** SHA-3's Keccak function divided into Theta, Rho, Chi, Pi and Iota blocks [90].

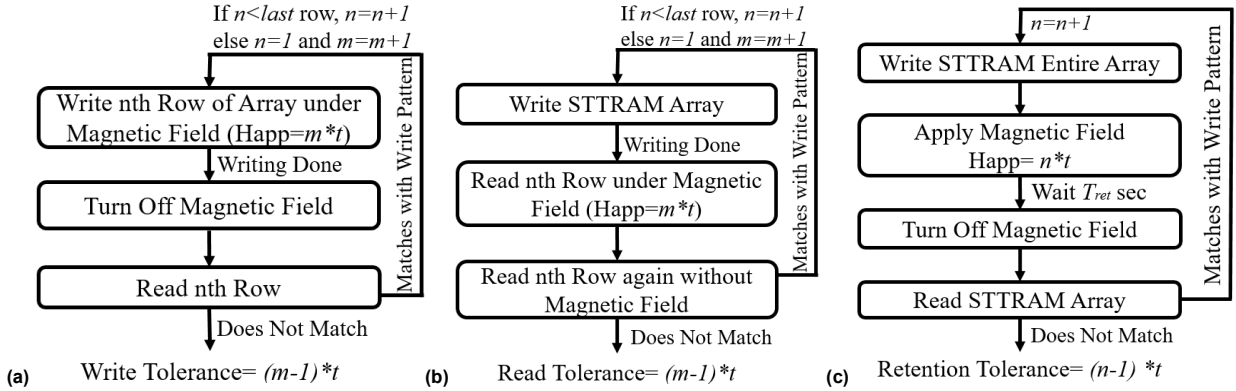

**FIGURE 26.** Algorithm for finding, (b) write tolerance; (c) read tolerance; and, (c) retention tolerance.

(v) In-Memory Acceleration of Classic McEliece Encoder (iMACE) [91]: The McEliece crypto-system based on the general decoding problem is one of the front runner candidates for post-quantum cryptography. However, the energy-efficiency is limited by the heavy data traffic between the processing elements and the memory. In memory-computing (IMC) architectures can remove the energy-efficiency barriers posed by Von-Neumann computing due to the movement of data between the processor and the memory. iMACE is an encoder designed using RRAM-based IMC. The work implements DCIM [87] architecture which is energy efficient. The work also incurs lower memory footprint compared to other implementation of McEliece crypto-systems.

(vi) STT-CiM [92]: In [92], an MRAM based computing in memory technique (STT-CiM) has been proposed. Unique characteristics of MRAM are leveraged to enable multiple wordlines within an array simultaneously. Therefore, data stored in multiple rows can be sensed from a single access. The authors also propose modifications to the peripheral circuits and enable logic/arithmetic and complex vector operations.

Several other techniques are proposed for IMC using STTRAM [93], SOT-MRAM [94-95] and PCM [96-98]. Since the data can be processed in the memory with lower energy, this leads to a significant improvement in the performance and bandwidth.

### 2) ATTACKS ON IMC
External fields such as magnetic field and temperature can bias the memory state to specific polarity. This will lead to erroneous data computation. Such polarity specific fault injection can be leveraged to extract secret information. Lower endurance of memories such as, PCM/RRAM can be leveraged to exhaust the memory cells and lead to DoS attacks. IMC using NVMs is susceptible to SCA. This is true since during read/write operation, the current drawn by the memory cells will show asymmetry. An adversary can launch the attack and extract the data being computed. Prior work has shown the vulnerability of CMOS-based computation to SCA. However, NVM-based IMC is more vulnerable since the asymmetry is significantly more compared to CMOS counterparts.

### 3) COUNTERMEASURES
(i) Physical shields can prevent tampering the IMC using external magnetic fields. Sensors can be implemented to detect a thermal or magnetic attack and discard the unreliable data during the attack. (ii) Noise injection during computation can obfuscate the data signature. This will prevent an SCA attack on NVM-based IMC.

## VIII. MEMORY TESTING TO DETECT VULNERABILITIES
In this section, some testing methodologies proposed in prior works are summarized which can capture potential security vulnerabilities after the NVM chips are manufactured.

### 1) SENSITIVITY TESTING
As mentioned before, spintronic memories store data in terms of the magnetic orientation of a ferromagnetic layer, and an external magnetic field can lower its retention [99] or even corrupt the data [58]. Therefore, spintronic memories can be tested for magnetic tolerance in all operating modes and rated them with their maximum tolerance. If a sensor detects an external magnetic field more than the rated tolerance, the information stored in the sensor can be discarded. Furthermore, all NVMs are susceptible to temperature. Temperature variation can cause read/write/retention failures. Therefore, all NVMs should be tested for thermal tolerance and rated accordingly.

MAGNETIC TOLERANCE TEST [100]:
Spintronic memories should be tested and certified during the write, read and retention mode separately since their tolerance for different modes could be different. Furthermore, chip to chip tolerance can identify the weakest chip due to process variation. Therefore, a tolerance test could discard chip which has a lower tolerance than the rated one. If a chip still incurs an attack more than the threshold value, a sensor can detect it and take necessary measures as proposed in [74, 87].

**Write Tolerance:** Write tolerance is the maximum magnetic field under which it can be written successfully at a specified write current with a specified write latency.

**Read Tolerance:** Read tolerance is the maximum magnetic field under which it can be read successfully without causing any disturbance to the bits at a specific read current with a specified read latency.

**Retention Tolerance:** Retention tolerance is the maximum external magnetic field under which it does not incur any data-corruption for a specified time period during retention mode. The algorithms to find the write tolerance, read tolerance and retention tolerance are shown in Fig. 26(a), 26(b) and 26(c), respectively [100]. Write tolerance depends on the data that is being written whereas, read and retention tolerance depends on the stored data. The worst-case write tolerance of a bit occurs when writing $0 \rightarrow 1$ since it incurs higher write time. The worst-case read tolerance and retention tolerance of a bit occur when the bit stores data '1'. The reason is data '0' (P state) is the preferred state for STTRAM i.e. writing $1 \rightarrow 0$ requires less write current and time. Therefore, all operating modes of a chip should be tested for their maximum tolerance.

THERMAL TOLERANCE TEST [101]:

Memory chips should be certified to operate successfully within a target temperature range (typical range: -10°C to 90°C):

(a) At high temperature, energy barrier between two states of the memory reduces. Therefore, the data retention time reduces and can lead to retention failure. Furthermore, read failure can occur since the reduction of resistance difference between two states leads to a reduction of sense margin and read disturb can occur since slight disturbance can flip the data at a lower energy barrier. Therefore, the manufacturer needs to test retention and read failure/disturb at high temperature.

(b) At low temperature, the energy barrier between the two states increases. Therefore, the read/write latencies increase and can lead to read/write failures.

The above discussion indicates that the manufacturer needs to certify the memory with a temperature range where the chip can operate successfully. If the memory incurs temperature out of that range, the stored data in the memory are no longer reliable.

Thermal tolerance test can be done using the algorithms proposed for the magnetic tolerance test (Section VIII.1) by applying an external thermal field instead of the magnetic field. This test can be combined with standard hot-cold test. The highest temperature at which the memory read failure/disturb does not occur and retention time meets the minimum target specification is the upper limit of thermal tolerance. The lowest temperature at which read/write operation does not fail is the lower limit of thermal tolerance.

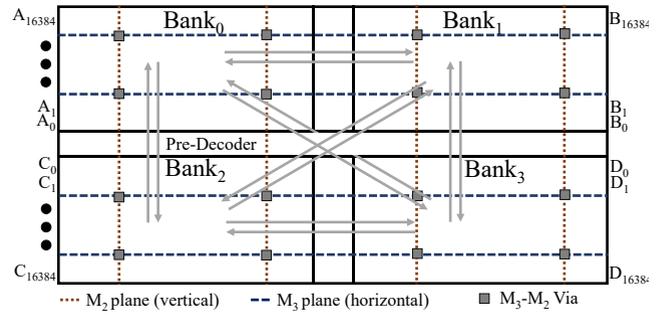

**FIGURE 27.** 4MB LLC diagram showing addresses of Bank0/Bank1/ Bank2/Bank3 as A/B/C/D from 0 to 16K. The grey arrows show different cases for testing the impact of supply noise.

2) SUPPLY NOISE TESTING [102-103]

An adversary can leverage supply noise to launch fault injection, DoS, information leakage and row hammer attack as discussed in Section III.2, III.3 and III.4. This is especially true if parallel operation are done to near-by independent memory banks and they have a strong coupling (lower resistance and capacitance between two address of two independent banks). Mostly, the victim bitcells are the weaker cells due to process variation.

In [102], a supply noise test technique is proposed to capture the impact of parallel write operations on the weaker bitcells. The method is further improved in [103] which implements test time compression technique leveraging unique data patterns. The work also divides the testing scenarios into 3 cases as shown in Fig. 27: i) accesses in adjacent banks (e.g. Bank0-Bank1); (ii) accesses in physically confronting banks (e.g. Bank0-Bank2); and (iii) accesses in diagonal banks (e.g. Bank0-Bank3). This cases successfully captures a weaker bitcells which gets affected by near-by parallel operations. Once identified, either parallel accesses can be restricted to such weak cells or the chips with weaker bitcells can be discarded.

3) ENDURANCE TESTING [101]

NVM performance can degrade over time due to physical breakdown (STTRAM/RRAM/PCM) or resistance drift (RRAM/PCM). STTRAM/MRAM have an oxide layer in their storage element, MTJ, and RRAM has oxide layer between two electrodes in its bitcell. Oxide might breakdown due to high Iwrite leading to function failure. It has also been reported that LRS changes 2X-10X and HRS changes 5X-100X in $TaO_2$ based RRAM due to variation. In PCM, time-dependent resistance drift in amorphous chalcogenide material is one of the major reliability concerns. Therefore, row hammer attack on NVM can a big security concern.

In [101], authors have proposed a novel test technique which can measure the endurance of the memory cell in a very short test time. The basic idea is to create the model of the physical parameters that changes as the cell is written multiple times. For example, Fig 28 (a) shows the change of ratio of resistance in HRS and LRS of RRAM bitcell and Fig. 28(b) shows the change of STTRAM bitcell resistance with respect to the number of times they are hammered. Once such modelling are done, the bitcells of a chip can be hammered using a DFT circuits (Fig. 29(a)). The corresponding waveform of the DFT circuit is shown in Fig. 29(b). The resistance of the bitcells can be measured and the effective endurance can be calculated by leveraging the relation shown in Fig. 29 and applying extrapolation. If the endurance is lower than the target or threshold endurance, they can be considered as vulnerable chips. Such chips can be discarded or can be implemented with proper mitigation techniques or in those application which does not require high endurance.

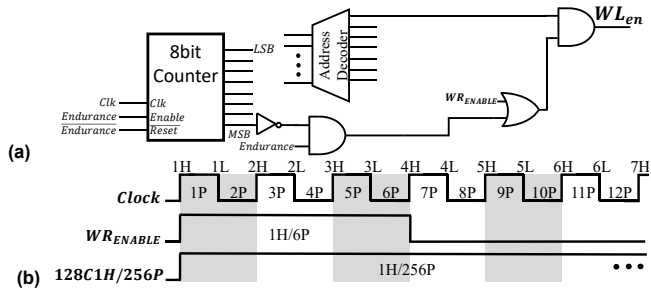

(a)

(b)

**FIGURE 28.** (a) DFT circuit for endurance test; (b) inputs waveforms of the proposed DFT circuit (Fig. 11(a)).

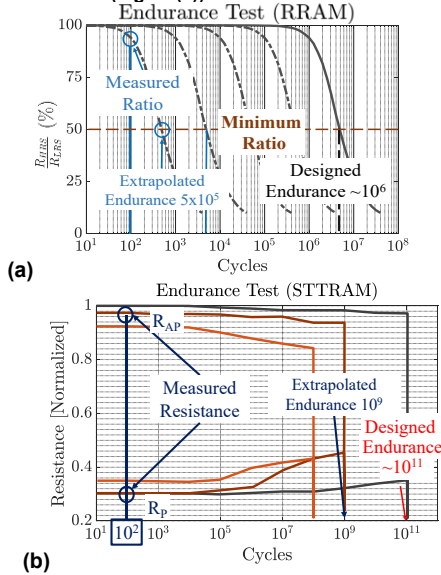

(a)

(b)

**FIGURE 29.** Endurance test for, (a) STTRAM; and, (b) RRAM using extrapolation for shorter test time.

## IX. FUTURE RESEARCH DIRECTION

Emerging NVMs carry significant potential to bridge performance, energy-efficiency and density gaps in existing and emerging systems. In intermittently powered systems, NVMs can store processor state whereas in low-power and high-performance systems NVMs can replace existing memories or provide additional level of cache. In this paper, we summarized the new threat surface considering the examples of several important vulnerabilities, attack vectors and potential countermeasures associated with NVMs. Next generation systems are already adopting some of these memories making the threats very real. Our analysis suggested that adversaries can exploit the complex device properties to break cryptography and/or launch system security attacks. The device designers, cryptography community and computer scientists need to work together to understand the new threat surface and develop countermeasures. Some important focus areas of future research should include:

**New sources of vulnerabilities:** Although the vulnerabilities originating from the devices themselves should be examined further, the community needs to think beyond that to identify new sources. For examples, NVMs employ several peripherals e.g., assist mechanisms that can act as potential sources of fault injection. Such vulnerabilities have been explored for SRAM and DRAM. NVMs can also employ system level features e.g., wear-leveling that can be turned against the NVMs/disabled to launch attacks. These issues need to be thoroughly investigated.

**Attacks on storage NVMs:** The security and design community needs to consider new attack vectors on the NVMs beyond DoS and fault injection such as, Trojan-induced attacks. Some topics require more efforts e.g., information leakage since multiple directions exist to exploit NVM features, peripherals and system-level mechanisms to leak data. So far, only external field-based tampering is investigated.

**Attacks on compute NVMs:** Variety of NVMs are being explored as alternative substrate for computation. The compute memories are different than the storage memories since their mode of operation reaches beyond simple storage. The attacks on these memories need to be understood to develop countermeasures.

**NVM enabled attacks:** Once NVMs are integrated in the system, they can be exploited by adversaries to launch new sources of attack. One example illustrated in this paper in Trojan design however, scopes of other attacks are equally likely and need further exploration.

**NVM based Trojan triggers:** Once NVMs are integrated in the system, they can be exploited by hardware Trojans. More research is needed to identify other NVM properties that can be exploited for Trojan trigger and payload design. NVM-based Trojans can be dangerous since system level techniques fails to detect them. This mandates further research for detection and prevention of NVM Trojans.

**Attack detection/sensing and prevention:** One of the important aspect of countering the attacks is to detect them. We showed an example of magnetic sensor to detect attacks however, such principles need to be extended to other attacks as well. Prevention will require elimination of vulnerabilities through device engineering and/or by employing new low-overhead techniques. Further research is required in these directions to secure the NVMs.

NVM testing: New testing techniques can be investigated to detect NVM vulnerabilities after manufacturing. For example, new testing techniques can be investigated that can identify the chips with hardware Trojans successfully at lower overhead and discard them.

## X. CONCLUSION

In this work, we summarized the basics of emerging NVMs and their vulnerabilities. We discussed the privacy and security issues for NVM-based cache and main memory. We also summarized various state-of-the-art countermeasures. We present a discussion on the vulnerability of NVM-based IMC and on the premise of employing hardware Trojan leveraging NVM. We also described test techniques to capture some of the security and privacy issues of NVM chips after manufacturing in short test time. Finally, we presented a discussion on the future topic of research on security/privacy of NVMs.

## ACKNOWLEDGEMENTS

The authors acknowledge the support from Anupam Chattopadhyay, Shivam Bhasin, Sumeet. Gupta, Jongsung Park, Rashmi Jha, Sandeep Thirumala, Alex Jones, Alex Yuan, Anirudh Iyengar, Seyedhamidreza Motaman, Asmit De, Rekha Govindaraj, Karthikeyan Nagarajan, Sina Sayyah Ensan and Nitin Rathi. The authors also acknowledge the financial support by Semiconductor Research Corporation (SRC) (2847.001), National Science Foundation (NSF) (CNS-1722557, CCF-1718474, DGE-1723687 and DGE-1821766) and DARPA Young Faculty Award (D15AP00089).

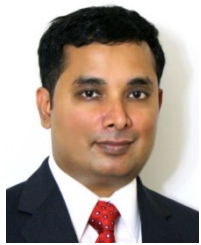

**Swaroop Ghosh** (S'04, SM'13) received his Ph.D. from Purdue in 2008. He is an Assistant Professor at Penn State. Dr. Ghosh was senior research and development engineer in Advanced Design, Intel Corp from 2008 to 2012. His research interests include low-power circuit design and hardware security. He is a senior member of IEEE and NAI.

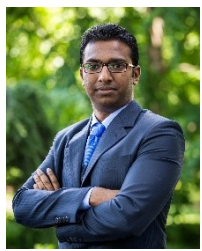

**Mohammad Nasim Imtiaz Khan** (S'17) received his Ph.D. from The Pennsylvania State University (Penn State) in 2019. Dr. Khan received his bachelor's from the department of Electrical Engineering of Bangladesh University of Engineering and Technology (BUET), 2014 and currently employed at Intel as SSD System Media Engineer. His research interests include hardware security and security and privacy of emerging NVM.